\documentclass[11pt]{article}
\usepackage{amsmath}
\usepackage{amsfonts}

\usepackage{amsmath,amssymb}
\usepackage{graphicx}
\usepackage{color}
\usepackage{url}
\usepackage{float}

\renewcommand{\arraystretch}{1.3}

\newtheorem{theorem}{Theorem}
\newtheorem{lemma}{Lemma}
\newtheorem{corollary}{Corollary}

\newtheorem{definition}{Definition}
\newtheorem{proposition}{Proposition}
\newtheorem{remark}{Remark}

\newcommand{\RNum}[1]{\uppercase\expandafter{\romannumeral #1\relax}}

\newcommand{\gf}{{\mathbb{F}}}

\newcommand{\ls}[1]
    {\dimen0=\fontdimen6\the\font\lineskip=#1\dimen0
     \advance\lineskip.5\fontdimen5\the\font
     \advance\lineskip-\dimen0
     \lineskiplimit=0.9\lineskip
     \baselineskip=\lineskip
     \advance\baselineskip\dimen0
     \normallineskip\lineskip\normallineskiplimit\lineskiplimit
     \normalbaselineskip\baselineskip
     \ignorespaces}

\begin{document}

\bibliographystyle{abbrv}

\title{Further Investigation on Differential Properties of the Generalized Ness-Helleseth Function}

\author{Yongbo Xia\thanks{Y. Xia is with the School of Mathematics and Statistics, and 
  also with the Hubei Key Laboratory of Intelligent Wireless Communications,
  South-Central Minzu University, Wuhan 430074, China (xia@mail.scuec.edu.cn).},  Chunlei Li
\thanks{C. Li  and T. Helleseth are with the  Department of Informatics, University of
Bergen, N-5020 Bergen, Norway (e-mail: chunlei.li@uib.no;
tor.helleseth@uib.no).}, Furong Bao\thanks{F. Bao is with the School of Mathematics, Southwest Jiaotong University, Chengdu 610031, China (baokekebpsp@163.com).}, Shaoping Chen\thanks{S. Chen is
  with the Hubei Key Laboratory of Intelligent Wireless Communications,
  South-Central Minzu University, Wuhan 430074, China (spchen@scuec.edu.cn).},
 and Tor Helleseth \footnotemark[2]}
\date{}
\maketitle

\thispagestyle{plain} \setcounter{page}{1}

\begin{abstract}
Let $n$ be an odd positive integer, $p$ be a prime with $p\equiv3\pmod4$, $d_{1} = {{p^{n}-1}\over {2}} -1 $ and $d_{2} =p^{n}-2$. The function defined by $f_u(x)=ux^{d_{1}}+x^{d_{2}}$ is called the generalized Ness-Helleseth function over $\mathbb{F}_{p^n}$, where $u\in\mathbb{F}_{p^n}$.  It was  initially studied by Ness and Helleseth in the ternary case.  In this paper, for $p^n \equiv 3 \pmod 4$ and $p^n \ge7$, we provide the necessary and sufficient condition for $f_u(x)$ to be an APN function. In addition,  for each $u$ satisfying $\chi(u+1) = \chi(u-1)$, the differential spectrum of $f_u(x)$ is investigated, and it is expressed in terms of some quadratic character sums of cubic polynomials, where $\chi(\cdot)$ denotes the quadratic character of $\gf_{p^n}$.

\vspace{2mm}


\end{abstract}

\ls{1.5}

\section{Introduction}\label{sec-intro}

Let $\mathbb{F}_{p^n}$ be the finite field with $p^n$ elements and $\mathbb{F}_{p^n}^*=\gf_{p^n}\setminus \{0\}$, where $p$ is a prime number and $n$ is a positive integer.  Let $F(x)$ be a function from  $\mathbb{F}_{p^n}$ to itself. The \textit{derivative function} of $F(x)$ at an element $a$ in $\gf_{p^n}$, denoted by $\mathbb{D}_aF(x)$,  is given  by
$$\mathbb{D}_aF(x)=F(x+a)-F(x).$$
For any $a,\,b \in \gf_{p^n}$,  let  $\delta_F(a,b)=|\{x \in \gf_{p^n}~| ~\mathbb{D}_{a}F(x)=b\}|,$
where $|S|$ denotes the cardinality of a set $S$.
The \emph{differential uniformity} of $F(x)$ is defined as
\[\delta(F)=\max \{ \delta_F(a,b)~| ~a \in \mathbb{F}_{p^n}^*,\,\, b \in \mathbb{F}_{p^n}\}.\]
A function $F(x)$ is said to be differentially $\delta$-uniform if  $\delta(F)=\delta$. Differential uniformity is an important concept in cryptography introduced by Nyberg \cite{Nyberg1994}, which can be used to quantify the security of the block cipher with respect to the differential attacks if $F(x)$ is used as an S-box in the cipher. The lower the differential uniformity of $F(x)$, the better is its resistance against a differential attack. In particular, when $\delta(F)=1$, $F(x)$ is said to be a perfect nonlinear (PN) function; when $\delta(F)=2$, $F(x)$ is known as an almost perfect nonlinear (APN) function. During the last $30$ years, PN and APN functions have been thoroughly investigated.  They are of importance in cryptography \cite{Nyberg1991,Nyberg1994}, and  also useful  in coding theory \cite{YCD2006,WLZ2020,XTD2022} and mathematics \cite{DO1968,Coulter97,Coulter2008,Ganley1975,Dy2006}. Recent research on PN and APN functions can be found in \cite{ Leander2022,Faruk2023, Faruk2022} and the references therein.

Compared with the differential uniformity, the differential spectrum of a nonlinear function reflects more information about its differential property. Its definition is given as follows.

\begin{definition}\label{def1}
Let $F(x)$ be a function over $\gf_{p^n}$ with differential uniformity $\delta$, and define
$$\omega_i=|\{ (a,b)\in \mathbb{F}_{p^n}^*\times \mathbb{F}_{p^n} \mid \delta_{F}(a,b)=i\}|,~0\le i \le \delta.$$
The differential spectrum of $F(x)$ is defined to be
an ordered sequence
\[
\mathbb{S} = [\omega_0, \omega_1, \ldots, \omega_{\delta}].
\]
\end{definition}
According to Definition \ref{def1}, the differential spectrum $\mathbb{S}$ satisfies the following two identities:
\begin{equation}\label{prop}
\sum\limits_{i=0}^\delta \omega_i=(p^n-1)p^n\,\,{\rm and}\,\,\sum\limits_{i=0}^\delta \left(i\times \omega_i\right)=(p^n-1)p^n,
\end{equation}
where the first identity stems from the number of pairs $(a,b)\in \gf_{p^n}^*\times \gf_{p^n}$ and the second one comes from the number of pairs $(a,x)\in \gf_{p^n}^*\times \gf_{p^n}$. These two identities are useful in calculating the differential spectrum of $F(x)$.

 Note that when $F(x)$ is  a power function, i.e.,  $F(x)=x^d$ for some positive integer $d$,  one can easily see that $\delta_F(a,b)=\delta_F(1,{b/{a^d}})$  for all $a\in \mathbb{F}_{p^n}^*$ and $b\in \mathbb{F}_{p^n}$. Thus, the differential spectrum of a power function $f(x)=x^d$ can be simplified as
$\overline{\mathbb{S}} = [\overline{\omega}_0, \overline{\omega}_1, \ldots, \overline{\omega}_{\delta}]$, where $$\overline{\omega}_i=|\left\{b\in \mathbb{F}_{p^n}\mid \delta_F(1, b)=i\right\}|,\,\,0\leq i\leq \delta.$$
It is an interesting topic to determine the differential spectra of cryptographic functions
with low differential uniformity. However, this problem is relatively challenging, and currently remains largely unexplored. Up to now, the known results are primarily concerned with power functions, with only a few studies addressing polynomials.  For more details, the reader is referred to the recent work \cite{Xia-Bao2023}.

Let $n$ be an odd positive integer, $p$ be an odd prime satisfying $p\equiv3\pmod4$, $d_{1} = {{p^{n}-1}\over {2}} -1 $ and $d_{2} =p^{n}-2$. Define
\begin{equation}\label{mainfun}
f_u(x)=ux^{d_{1}}+x^{d_{2}},
\end{equation}
where $u\in\mathbb{F}_{p^n}$. The function $f_u(x)$ was originally studied by Ness and Helleseth in the ternary case \cite{Ness-Helleseth-2007}, and was further investigated in \cite{Xia-Bao2023}, where it was called \textit{ the ternary Ness-Helleseth function}. For general odd primes $p$, we call $f_u(x)$  \textit{the generalized Ness-Helleseth function} in this paper. For $p^n \equiv 3 \pmod 4$ and $p^n \ge7$, the differential properties of  the generalized Ness-Helleseth function $f_u(x)$ were partly studied by Zeng et al. in \cite{Zeng2007} and by Zha in his Phd dissertation \cite{Zha2008}. Let $\chi(\cdot)$ denote the quadratic character of $\mathbb{F}_{p^n}$ which is defined by
\begin{equation*}\chi(x)=\left\{
\begin{array}{ll}
1, &~ \mathrm{if}~x~\mathrm{is~a~square~in}~\mathbb{F}_{p^n}^*,\\
-1, &~ \mathrm{if}~x~\mathrm{is~a~nonsquare~in}~\mathbb{F}_{p^n}^*,\\
0, &~ \mathrm{if}~x=0.
\end{array} \right.
\end{equation*} Their results are  summarized as follows.

\begin{theorem}\label{result of zeng}\cite{Zeng2007}
Let $p^n \equiv3 {\pmod 4}$, $p^n\geq7$ and $u$ be an element in $\mathbb{F}_{p^n}$ such that $\chi(u+1)=\chi(u-1)=-\chi(5u+3)$ or $\chi(u+1)=\chi(u-1)=-\chi(5u-3)$. Then, the generalized Ness-Helleseth function $f_u(x)$  defined  in (\ref{mainfun}) is an APN function.
\end{theorem}

\begin{theorem}\label{result of zha}\cite{Zha2008}
Let  $p^n \equiv3 {\pmod 4}$. Then, the differential uniformity of the generalized Ness-Helleseth function  $f_u(x)$ defined in \eqref{mainfun} is equal to $3$ when  $u$ satisfies $\chi(u+1)=\chi(u-1)=\chi(5u+3)=\chi(5u-3)$. Moreover, such an element $u$ exists if $p^n>19$.
\end{theorem}

For simplicity, we introduce the following sets:
\begin{equation}\label{setu}
\left \{\begin{array}{lll}
    \mathcal{U}_0 &=\{u\in\gf_{p^n} \mid \chi(u+1)\neq\chi(u-1)\},\\
    \mathcal{U}_1 &=\{u\in\gf_{p^n} \mid \chi(u+1)=\chi(u-1)\},\\
    \mathcal{U}_{10}&=\{u\in\gf_{p^n} \mid \chi(u+1)=\chi(u-1)=-\chi(5u+3)\},\\
      \mathcal{U}_{11}&=\{u\in\gf_{p^n} \mid \chi(u+1)=\chi(u-1)=-\chi(5u-3)\},\\
    \mathcal{U}_{12} &=\{u\in\gf_{p^n} \mid \chi(u+1)=\chi(u-1)=\chi(5u+3)=\chi(5u-3)\}.\\
\end{array}\right.
\end{equation}
It can be verified that when $p^n \equiv3 {\pmod 4}$, we have $\gf_{p^n}=\mathcal{U}_0\cup\mathcal{U}_1$, $\mathcal{U}_0\cap\mathcal{U}_1=\emptyset$, $\{0,\pm1,\pm\frac{3}{5}\}\subseteq\mathcal{U}_0$,
$\mathcal{U}_{1}=\left(\mathcal{U}_{10}\cup\mathcal{U}_{11}\right)\cup\mathcal{U}_{12}$ and $\left(\mathcal{U}_{10}\cup\mathcal{U}_{11}\right)\cap\mathcal{U}_{12}=\emptyset$. Moreover, we have that $|\mathcal{U}_1|=\frac{p^n-3}{2}$,  $|\mathcal{U}_{10}|=|\mathcal{U}_{11}|$, $|\mathcal{U}_{10}\cup \mathcal{U}_{11}|=\frac{p^n-1+2\chi(5)}{4}$ and $ |\mathcal{U}_{12}|>0$ if $p^n>19$ (see Remark \ref{remark1}).

In the present paper, we continue to investigate the differential properties of the generalized Ness-Helleseth function under the condition that $p^n \equiv 3 \pmod 4$ and $p^n \ge7$. Note that the condition $p^n\ge 7$ means that when $p=3$, we should add the condition $n\ge 3$. This condition ensures $|\mathcal{U}_{10}\cup \mathcal{U}_{11}|>0$. Similarly,  $p^n>19$ indicates that we require the condition $n\ge 3$ when $p=3$ or $7$. When $u=0$ ($0\in \mathcal{U}_0$), $f_u(x)$ is exactly the power mapping $x^{p^n-2}$. The differential spectrum of $x^{p^n-2}$ was completely determined in \cite{Zhang2020}.

\begin{theorem}\label{resultofu=0}\cite{HRS1999IT,DMMPW2003,Zhang2020}
	Let $p$ be an odd prime and $f(x)=x^{p^n-2}$ be the power function over $\gf_{p^n}$. Then,  the differential uniformity $\delta(f)$ of $f(x)$ is given as follows
	\begin{equation*}\label{du-of-fu}
		\delta(f)=\left \{\begin{array}{cl}
			3,&~ \mathrm{if}~p=3,\\
			2,&~ \mathrm{if}~p^n\equiv 2~~(\mathrm{mod}~3),\\
		    4,&~ \mathrm{if}~p^n\equiv 1~~(\mathrm{mod}~3).\\
			\end{array}\right.\end{equation*}
Moreover, the differential spectrum of $f(x)$ is given by

\noindent (i) $\left[\,\overline{\omega}_0=\frac{3^n-1}{2},\, \overline{\omega}_1=0,\overline{\omega}_2=\frac{3^n-3}{2},\,\overline{\omega}_3=1\,\right]$ if $p=3$;

\noindent (ii) $\left[\,\overline{\omega}_0=\frac{p^n-1}{2},\, \overline{\omega}_1=1,\,\overline{\omega}_2=\frac{p^n-1}{2}\,\right]$ if $p^n\equiv 2 \pmod 3$;

\noindent (iii) $\left[\,\overline{\omega}_0=\frac{p^n+1}{2},\,\overline{\omega}_1=1,\,\overline{\omega}_2=\frac{p^n-5}{2},\,\overline{\omega}_3=0,\overline{\omega}_4=1\,\right]$ if $p^n\equiv 1 \pmod 3$.
\end{theorem}

Based on these known results, in order to completely determine the differential uniformity of the generalized Ness-Helleseth function $f_u(x)$, it remains to calculate the differential uniformity of $f_u(x)$ for $u\in \mathcal{U}_0\setminus\{0\}$. This problem has been solved for $p=3$ in \cite{Xia-Bao2023}.  In this paper, we will deal with this problem for general primes satisfying $p^n\equiv3\pmod 4$.

The rest of this paper is organized as follows. In Section \ref{pre}, we introduce some auxiliary results which will be used in the subsequent sections. In Section \ref{main-resultforFp-1}, by investigating the derivative equation of $f_u(x)$ and utilizing the theory of quadratic character sums, we determine the differential uniformity of $f_u(x)$ for $u\in \mathcal{U}_0\setminus \{0\}$. As a byproduct,  the differential spectrum of $f_u(x)$ for $u=\pm1$ is also calculated. In Section \ref{main-resultforFp-2}, for each $u\in \mathcal{U}_1$, the differential spectrum of $f_u(x)$ is studied and expressed in terms of some quadratic character sums of cubic polynomials. Section \ref{con-remarks} concludes our study.

\section{Preliminaries}\label{pre}

Throughout this paper, we always assume that $p$ is an odd prime. It is well-known that $\chi(-1)=-1$ when $p^n\equiv3\pmod 4$. For a square element $s\in \mathbb{F}_{p^n}$, we use $\pm \sqrt{s}$ to denote the two square roots of $s$.  Let $\mathbb{F}_{p^n}[x]$ denote the polynomial ring over $\gf_{p^n}$. We recall some results about the quadratic character sums of the form
$\sum\limits_{x\in\mathbb{F}_{p^n}}\chi(f(x))$ with $f(x)\in\mathbb{F}_{p^n}[x]$.
\begin{lemma} \cite[Theorem 5.48]{FF}\label{charactersumquadratic} Let $f(x)=a_2x^2+a_1x+a_0\in\gf_{p^n}[x]$ with $p$ odd and $a_2\neq0$. Put $d=a^2_1-4a_0a_2$ and let $\chi(\cdot)$ be the quadratic character of  $\mathbb{F}_{p^n}$. Then
\begin{eqnarray*}
\sum_{x\in\mathbb{F}_{p^n}}\chi(f(x))=\left\{
\begin{array}{lllll}
-\chi(a_2), ~~~~~~~\mathrm{if}~d\neq0,\\
(p^n-1)\chi(a_2), ~\mathrm{if}~d=0.
\end{array} \right.\ \
\end{eqnarray*}
\end{lemma}

Assume that $f(x)$ is a cubic polynomial in $\gf_{p^n}[x]$ with distinct roots in its splitting field.
Let $E(\gf_{p^n})$ be the elliptic curve over $\gf_{p^n}$ defined by
\begin{equation*}
	E(\gf_{p^n}): y^2=f(x),
\end{equation*}
and $|E(\mathbb{F}_{p^n})|$ denote the number of $\mathbb{F}_{p^n}$-rational
points (with the extra point at infinity) on the curve $E(\gf_{p^n})$. By Application 1.3 in \cite[P. 139, Chap. V]{SilveEC}, we have
\begin{equation*}\label{ecn}
|E(\mathbb{F}_{p^n})|=p^n+1+\sum\limits_{x\in\mathbb{F}_{p^n}}\chi(f(x)).
\end{equation*}
The above relation allows for calculating quadratic character sum of some $f(x)$ for which the number of $\mathbb{F}_{p^n}$-rational points on the curve  $E(\mathbb{F}_{p^n})$ is known.

The following theorem
 provides lower and upper bounds for the character sum $\sum\limits_{x\in\mathbb{F}_{p^n}}\chi(f(x))$.

\begin{theorem}\cite[Theorem 5.41]{FF} \label{weilbound}Let $\psi(\cdot)$ be a multiplicative character of $\mathbb{F}_{p^n}$ of order $m>1$ and let $f\in \mathbb{F}_{p^n}[x]$ be a monic polynomial of positive
 degree that is not an $m$th power of a polynomial. Let $d$ be the number of distinct roots of $f$ in its splitting field over $\mathbb{F}_{p^n}$. Then for every $a\in \mathbb{F}_{p^n}$ we have
 $$ \left| \sum\limits_{x\in\mathbb{F}_{p^n}}\psi(af(x))\right|
 \leq (d-1)\sqrt{p^n}.$$

\end{theorem}

Let $\mathcal{C}_0$ and $\mathcal{C}_1$ denote the sets of squares and nonsquares in $\mathbb{F}_q^*$, respectively.
 For $0\le i, j\le 1$, the cyclotomic number $(i,j)$ is defined as $(i,j)=|\{(x_i,x_j)\in \mathcal{C}_i\times \mathcal{C}_j\,|\,x_i+1=x_j\}|$. The following results
 about the cyclotomic numbers  $(i,j)$ are well-known.

\begin{lemma}\label{lem cyclotomic}(\cite[Lemma 6]{Cyclotomy1967}) The cyclotomic numbers $(i,j)$
are given as follows:

\noindent(\textrm{i}) if $p^n \equiv 1\,\,({\rm mod}\,\,4)$, then
\begin{equation*}
(0,0)=\frac{p^n-5}{4},\,\, (0,1)=(1,0)=(1,1)=\frac{p^n-1}{4};
\end{equation*}
\noindent(\textrm{ii}) if $p^n \equiv 3\,\,({\rm mod}\,\,4)$, then
\begin{equation*}
(0,0)=(1,0)=(1,1)=\frac{p^n-3}{4},\,\, (0,1)=\frac{p^n+1}{4}.
\end{equation*}
\end{lemma}

\begin{remark}\label{remark1} Let $\mathcal{U}_1$ and $\mathcal{U}_{1i}$ $(i=0,1,2)$ be the sets defined in \eqref{setu}. Using Lemma \ref{lem cyclotomic}, we can show that $|\mathcal{U}_1|=\frac{p^n-3}{2}$. By Lemma \ref{charactersumquadratic}, we can prove $|\mathcal{U}_{10}\cup \mathcal{U}_{11} |=\frac{p^n-1+2\chi(5)}{4}$.
By Lemma \ref{charactersumquadratic} and Theorem \ref{weilbound}, we can derive that  $|\mathcal{U}_{12}|>0$ when $p^n>19$. This has been proved in \cite{Zha2008}.
\end{remark}

\section{The differential uniformity of $f_u(x)$ when $u\in\mathcal{U}_0\setminus\{0\}$}\label{main-resultforFp-1}
 In this section, we will determine the differential uniformity of $f_u(x)$ for each $u\in\mathcal{U}_0\setminus \{0\}$. First we investigate the derivative equation of $f_u(x)$:
\begin{equation}\label{diff-equ}
	\mathbb{D}_af_u(x)=u(x+a)^{\frac{p^n-1}{2}-1}+(x+a)^{p^n-2}-ux^{\frac{p^n-1}{2}-1}-x^{p^n-2}=b,
\end{equation}
where $(a,b)\in\mathbb{F}_{p^n}^*\times \mathbb{F}_{p^n}$. Let $N_u(a,b)$, $N_{u,1}(a,b)$ and $N_{u,2}(a,b)$ denote the number of solutions of \eqref{diff-equ} in $\mathbb{F}_{p^n}$, $\{0,-a\}$ and $\mathbb{F}_{p^n}\setminus\{0,-a\}$, respectively. Then, we have
\begin{equation}\label{N}
	N_u(a,b)=N_{u,1}(a,b)+N_{u,2}(a,b).
\end{equation}

In what follows, for each $(a,b)\in\mathbb{F}_{p^n}^*\times \mathbb{F}_{p^n}$ and $u\in\mathbb{F}_{p^n}$,  we will determine the values of $N_{u,1}(a,b)$ and $N_{u,2}(a,b)$.

By taking $x=0$ and $x=-a$ into the derivative equation  \eqref{diff-equ}, we have
\begin{equation}\label{N1}N_{u,1}(a,b)=\left\{
\begin{array}{ll}
2, &~ \mathrm{if}~u=0~\mathrm{and}~b=a^{-1},\\
1, &~ \mathrm{if}~u\neq0~\mathrm{and}~b=a^{-1}(1\pm u\chi(a)),\\
0, &~ \mathrm{otherwise}.
\end{array} \right.
\end{equation}
When $x\notin\{0,-a\}$, the derivative equation \eqref{diff-equ} can be reduced into
\begin{equation}\label{diff-equ2}
bx^2+(ab+u\tau_{0}-u\tau_{a})x+a(u\tau_{0}+1)=0,
\end{equation}
where $\tau_{a}=\chi(x+a)$ and $\tau_{0}=\chi(x)$. Following the same methodology as we did in our previous work \cite{Xia-Bao2023}, the derivative equation \eqref{diff-equ2} can  be reduced to four cases  in Table \ref{newtable0} for $b=0$ and to four quadratic equations in Table \ref{newtable1}
for $b\neq 0$. In each table, we present necessary information about the reduced equations, including their solutions $x$ in $\mathbb{F}_{p^n}$ and the associated values $x+a$.
In Table \ref{newtable1}, each reduced equation is quadratic, and the product of the two solutions $x_1$, $x_2$, denoted by $x_1x_2$, is also listed.
Notice that the solution(s) listed in each case of  Tables \ref{newtable0} and \ref{newtable1}
are not guaranteed to satisfy the corresponding condition on $(\tau_a, \tau_0)$.
This means that the reduced equations in Tables \ref{newtable0} and \ref{newtable1} may have
extraneous solutions with respect to \eqref{diff-equ2}.
	In the following, a solution of certain reduced equation in Tables \ref{newtable0}  or \ref{newtable1}  is termed
	\textit{\textbf{a desired solution}} if it indeed satisfies \eqref{diff-equ2}, or equivalently, if it satisfies the corresponding condition on $(\tau_a, \tau_0)$.

\begin{table*}[htb!]
	\caption{Four reduced equations of \eqref{diff-equ2} for $b=0$
	}\label{newtable0}
	\centering
	\setlength{\tabcolsep}{2mm}
	%		\resizebox{\linewidth}{!}{
		\begin{tabular}{|c|c|c|c|c|}
			\hline
			Case & \uppercase\expandafter{\romannumeral1} & \uppercase\expandafter{\romannumeral2} & \uppercase\expandafter{\romannumeral3} & \uppercase\expandafter{\romannumeral4}\\  \hline
			$(\tau_a,\tau_0)$ & $(1,1)$ & $(1,-1)$ & $(-1,1)$ &$(-1,-1)$\\  \hline
			Equation & $a(u+1)=0$ & $
			2ux+a(u-1)=0$ & $2ux+a(u+1)=0$ &$a(u-1)=0$\\ \hline
			$x$ & - & $-\frac{a(u-1)}{2u}\,(u\neq 0)$ & $-\frac{a(u+1)}{2u}\,(u\neq 0)$ &  - \\ \hline
			$x+a$ & - & $\frac{a(u+1)}{2u}\,(u\neq 0)$ & $\frac{a(u-1)}{2u}\,(u\neq 0)$ &  - \\ \hline
		\end{tabular}
		%	}
\end{table*}

\begin{table*}[htb!]
	\renewcommand{\arraystretch}{2.2}
	\caption{Four reduced equations of \eqref{diff-equ2} for $b\neq 0$ }\label{newtable1}
	\centering
	\setlength{\tabcolsep}{1.5mm}
	\resizebox{\linewidth}{!}
	{
		\begin{tabular}{|c|c|c|c|c|}
			\hline
			{ Case} & {\uppercase\expandafter{\romannumeral1}} & {\uppercase\expandafter{\romannumeral2} }& {\uppercase\expandafter{\romannumeral3} }& {\uppercase\expandafter{\romannumeral4}}\\  \hline
			{ $(\tau_a,\tau_0)$} & {$(1,1)$} & {$(1,-1)$} & {$(-1,1)$} &{$(-1,-1)$}\\  \hline
			{ Equation} & $bx^2+bax+a(u+1)=0$ & $bx^2+(ab-2u)x-a(u-1)=0$ &$bx^2+(ab+2u)x+a(u+1)=0$ & $bx^2+bax-a(u-1)=0$ \\ \hline
			{ $x$} & $\frac{-ab\pm \sqrt{a^2b^2-4(u+1)ab}}{2b}$ & $\frac{2u-ab\pm \sqrt{a^2b^2-4ab+4u^2}}{2b}$ & $\frac{-2u-ab\mp\sqrt{a^2b^2-4ab+4u^2}}{2b}$ &$\frac{-ab\pm\sqrt{a^2b^2+4(u-1)ab}}{2b}$\\ \hline
			{ $x+a$} &$\frac{ab\pm \sqrt{a^2b^2-4(u+1)ab}}{2b}$& $\frac{2u+ab\pm \sqrt{a^2b^2-4ab+4u^2}}{2b}$ & $\frac{-2u+ab\mp\sqrt{a^2b^2-4ab+4u^2}}{2b}$ &$\frac{ab\pm\sqrt{a^2b^2+4(u-1)ab}}{2b}$\\ \hline
			{ $x_1x_2$} & $\frac{a(u+1)}{b}$ & $-\frac{a(u-1)}{b}$ & $\frac{a(u+1)}{b}$ &$-\frac{a(u-1)}{b}$\\  \hline
			{$x(x+a)$} & $-\frac{a(u+1)}{b}$ & $\frac{2u^2-ab\pm u\sqrt{a^2b^2-4ab+4u^2}}{b^2}$ & $\frac{2u^2-ab\pm u\sqrt{a^2b^2-4ab+u^2}}{b^2}$ & $\frac{a(u-1)}{b}$\\  \hline
		\end{tabular}
	}
\end{table*}

According to the information in Tables \ref{newtable0} and \ref{newtable1}, we can get the following results: Lemmas \ref{lemforNa0}, \ref{I-IV} and \ref{II-III}. Their proofs are similar to those of Lemmas 3, 4, 5 and 6 in \cite{Xia-Bao2023}, and thus we omit them here.

\begin{lemma}\label{lemforNa0} With the notation introduced above and the information in Table \ref{newtable0}, for any $a\in \mathbb{F}_{p^n}^*$ we have
	\begin{equation}\label{lem3fuNa0}
		N_{u,2}(a,0)=\left\{
		\begin{array}{ll}
			\frac{p^n-3}{4}, &~ \mathrm{if}~u\in \{\pm 1\},\\
			0, &~ \mathrm{if}~u\in \mathcal{U}_0\setminus \{\pm 1\},\\
			1, &~\mathrm{if}~u\in \mathcal{U}_1.
		\end{array} \right.
	\end{equation}
\end{lemma}

\begin{lemma}\label{I-IV} With the notation and information in Table \ref{newtable1}, we have
the following results:
	
\noindent (i) the reduced equation in Case I in Table \ref{newtable1}  contributes exactly one desired solution to \eqref{diff-equ2} if
\begin{equation*}\label{condi-for-one-solution-Case1}
	\chi\left(a^2b^2-4(u+1)ab\right)=1 ~ \mathrm{and}~ \chi\left(\frac{a(u+1)}{b}\right)=-1,
\end{equation*} and  no desired solution otherwise;

\noindent (ii)  the reduced equation in Case IV in Table \ref{newtable1}  contributes exactly one desired solution if
\begin{equation*}\label{condi-for-one-solution-Case4}
	\chi\left(a^2b^2+4(u-1)ab\right)=1 ~ \mathrm{and}~ \chi\left(\frac{a(u-1)}{b}\right)=1,
\end{equation*} and no desired solution otherwise.

\end{lemma}

\begin{lemma}\label{II-III} With the notation and information in Table \ref{newtable1}, we have the following results:

 \noindent (i) if $x$ is a solution in Case II, then $-(x+a)$ is a solution in Case III, and vice versa. Moreover, $x, -(x+a)$ cannot both be desired solutions, and thereby Cases II and III can contribute at most two desired solutions in total;

 \noindent (ii) when $u\in\mathcal{U}_1$,  the two reduced equations in Cases II and III  in Table \ref{newtable1} contribute exactly one desired solution if
	\begin{equation*}\label{U1Case2-3}
		\chi(a^2b^2-4ab+4u^2)=1,
	\end{equation*}
	and no desired solution otherwise.
\end{lemma}

The following lemma is very useful in proving the main result of this paper.

\begin{lemma}\label{II-IV} With the notation and information in Table \ref{newtable1}, we further have the following results:
\noindent (i) when $u\in\mathcal{U}_0\setminus \{\pm1\}$, the two reduced equations in Cases II and III  in Table \ref{newtable1} contribute two desired solutions if
\begin{equation}\label{condiforCase2-3both-have-desired-solution}
	\left \{\begin{array}{lll}
   \chi(a^2b^2-4ab+4u^2)=1,\\
	\chi\left(2u^2-ab\pm u\sqrt{a^2b^2-4ab+4u^2}\right)=-1,
	\end{array}\right.\end{equation}
exactly one desired solution if
\begin{equation}\label{condiforCase2-3onlyonesolution}
	\left \{\begin{array}{lll}
\chi\left(\frac{a(u+1)}{b}\right)=1,\\
   a^2b^2-4ab+4u^2=0,\\
	\chi\left(2u^2-ab\pm u\sqrt{a^2b^2-4ab+4u^2}\right)=-1,
	\end{array}\right.\end{equation} and no desired solution otherwise;

\noindent (ii) when $u=\pm1$, the reduced equation in Case II (resp. III) contributes exactly one desired solution if
\begin{equation}\label{condi-for-one-solu|u|=1-0}
\begin{split}
\chi\left(4-2ab\right)=-1~ \mathrm{and}~ \chi\left(2b\right)=1
\left(\mathrm{resp.}~\chi\left(4-2ab\right)=-1~ \mathrm{and}~ \chi\left(2b\right)=-1\right),
\end{split}
\end{equation}
and no desired solution otherwise.
\end{lemma}
{\em Proof:} (i) When $u\in\mathcal{U}_0\setminus \{\pm1\}$, we have $\chi(u+1)=\chi(1-u)\in \{\pm 1\}$, which further implies $\chi\left(\frac{a(u+1)}{b}\right)=-\chi\left(\frac{a(u-1)}{b}\right)\in \{\pm 1\}$ for any $(a,b)\in \gf_{p^n}^*\times \gf_{p^n}^*$. Thus, in what follows we consider two separate cases:  $\chi\left(\frac{a(u+1)}{b}\right)=1$ and $\chi\left(\frac{a(u+1)}{b}\right)=-1$.

{\textit{Case 1}:} $\chi\left(\frac{a(u+1)}{b}\right)=-1$. Then $\chi\left(\frac{a(u+1)}{b}\right)=-\chi\left(\frac{a(u-1)}{b}\right)=-1$. In this case, the discriminant of the two reduced equations in Cases II and III is not equal to zero. Otherwise, we have $(ab-2u)^2=-4ab(u-1)$, which implies $\chi(-ab(u-1))=-\chi\left(\frac{a(u-1)}{b}\right)=1$, contradicting  the previous condition  $-\chi\left(\frac{a(u-1)}{b}\right)=\chi\left(\frac{a(u+1)}{b}\right)=-1$.  Thus, we only consider that case where $\chi(a^2b^2-4ab+4u^2)=1$. Then the reduced equations in Cases II and III both have two distinct solutions in $\mathbb{F}_{p^n}\setminus \{0,-a\}$. Let $x_1$ and $x_2$ be the two solutions of the reduced equation in Case II. According to Table \ref{newtable1},  \begin{equation}\label{prop2relation1}
	\chi(x_1x_2)=\chi\left(-\frac{a(u-1)}{b}\right)=-1.
\end{equation}  By Lemma \ref{II-III} (i),  $-(x_1+a)$ and  $-(x_2+a)$ are the solutions of the reduced equation in Case III. Then,

 \begin{equation}\label{prop2relation2}
 	\chi\left((x_1+a)(x_2+a)\right)=\chi\left(\frac{a(u+1)}{b}\right)=-1.
 \end{equation}
From \eqref{prop2relation1} and \eqref{prop2relation2}, we get
$\chi(x_1(x_1+a)x_2(x_2+a))=1$, which implies that

$$\chi(x_1(x_1+a))=\chi(x_2(x_2+a))=1$$ or $$\chi(x_1(x_1+a))=\chi(x_2(x_2+a))=-1.$$
If the former holds, then $\chi(x_i)=\chi(x_i+a)$, $i=1,2$, and thus the reduced equations in Cases II and III contribute no desired solution.   If the latter holds,  we can assume,  without loss of generality,  that $\chi(x_1)=-1$ and $\chi(x_2)=1$.  Then we have
\begin{equation*}\label{condi-case2-3-both-have-root}
\left \{\begin{aligned}
&\left(\chi(x_1),\chi(x_1+a)\right)=(-1,1), \\
&\left(\chi(x_2),\chi(x_2+a)\right)=(1,-1).
\end{aligned}\right.
\end{equation*}
According to the information in Table  \ref{newtable1}, one knows that $x_1$ is the desired solution from  Case II and  $-(x_2+a)$ is the desired solution from Case III.

The arguments above  show that when $u\in\mathcal{U}_0\setminus \{\pm 1\}$ and $\chi\left(\frac{a(u+1)}{b}\right)=-1$, Cases II and III can contribute two desired solutions in total (actually each of them contributes one desired solution) or no desired solution. When they contribute two desired solutions iff $\chi(a^2b^2-4ab+4u^2)=1$ and $\chi(x_1(x_1+a))=\chi(x_2(x_2+a))=-1$.
 By Table \ref{newtable1}, $\chi(x_1(x_1+a))=\chi(x_2(x_2+a))=-1$ is equivalent to  $\chi(2u^2-ab\pm u\sqrt{a^2b^2-4ab+4u^2})=-1$.  This yields the conditions \eqref{condiforCase2-3both-have-desired-solution} in this case.

{\textit{Case 2}:} $\chi\left(\frac{a(u+1)}{b}\right)=1$. Then $\chi\left(\frac{a(u+1)}{b}\right)=-\chi\left(\frac{a(u-1)}{b}\right)=1$.  If $\chi(a^2b^2-4ab+4u^2)=0$, then the reduced equations in Cases II and III both have a unique solution in $\mathbb{F}_{p^n}\setminus \{0,-a\}$. Let $x$ be the solution of Case II, then $-(x+a)$ is the unique solution of Case III. Note that $\chi(x(x+a))=1 $ or $-1$. We thereby consider two subcases. If $\chi(x(x+a))=1$, then Cases II and III contribute no desired solution. If $\chi(x(x+a))=-1$, we have $(\chi(x),\chi(x+a))=(-1,1)$ or $(\chi(x),\chi(x+a))=(1,-1)$. If the former holds, then $x$ is a desired solution of Case II; if the latter holds, then $-(x+a)$
is a desired solution of Case III. Thus, when $u\in\mathcal{U}_0\setminus \{\pm1\}$, $\chi\left(\frac{a(u+1)}{b}\right)=1$ and $\chi(a^2b^2-4ab+4u^2)=0$, the reduced equations in Cases II and III can contribute at most one desired solution in total, and when they contribute exactly one desired solution if $\chi(x(x+a))=-1$. According to Table \ref{newtable1}, this condition holds iff $\chi\left(2u^2-ab\pm u\sqrt{a^2b^2-4ab+4u^2}\right)=-1$. This leads to the conditions in \eqref{condiforCase2-3onlyonesolution}.

If $\chi(a^2b^2-4ab+4u^2)=1$, then the reduced equations in Cases II and III both have two distinct solutions in $\mathbb{F}_{p^n}\setminus \{0,-a\}$.  Let  $x_1$ and $x_2$ denote the two solutions of the reduced equation in Case II. In a way similar to Case 1, we have
$$\chi(x_1x_2)=1, \chi\left((x_1+a)(x_2+a)\right)=1 ~\mathrm{and} ~\chi\left(x_1x_2(x_1+a)(x_2+a)\right)=1.$$
Then, there are two possibilities:
\begin{equation}\label{possb1}
	\left \{\begin{array}{lll}
   \chi\left(x_1(x_1+a)\right)=1,\\
	\chi\left(x_2(x_2+a)\right)=1,
	\end{array}\right.\end{equation}
and
\begin{equation}\label{possb2}
	\left \{\begin{array}{lll}
   \chi\left(x_1(x_1+a)\right)=-1,\\
	\chi\left(x_2(x_2+a)\right)=-1.
	\end{array}\right.\end{equation}
If \eqref{possb1} happens, there is no desired solution from Cases II and III. Next we consider the case that \eqref{possb2} happens. Since $\chi(x_1x_2)=1$, we have $(\chi(x_1),\chi(x_2))=(1,1)$ or $(-1,-1)$. If $(\chi(x_1),\chi(x_2))=(-1,-1)$, then $x_1$ and $x_2$ are the desired solutions of Case II; if $(\chi(x_1),\chi(x_2))=(1,1)$, then $-(x_1+a)$ and $-(x_2+a)$ are the desired solutions of Case III. This means that
if \eqref{possb2} happens, Cases II and III can contribute exactly two desired solutions in total. To ensure \eqref{possb2} holds, it suffices to have $\chi\left(2u^2-ab\pm u\sqrt{a^2b^2-4ab+4u^2}\right)=-1$.

The arguments in Cases 1 and 2 show that when $u\in\mathcal{U}_0\setminus \{\pm1\}$, the two reduced equations in Cases II and III  of Table \ref{newtable1} contribute two desired solutions if \eqref{condiforCase2-3both-have-desired-solution} holds, one desired solution if \eqref{condiforCase2-3onlyonesolution} holds, and no desired solution otherwise.

(ii) First suppose that $u=1$. Then, the reduced equation in Case II becomes  $bx^2+(ab-2)x=0$, which has two solutions $x=0$ and $x=\frac{2-ab}{b}$. Only $x=\frac{2-ab}{b}$ can be a desired solution. According to Table  \ref{newtable1},  it is a desired
solution of Case II iff
$$\chi\left( \frac{2-ab}{b}\right)=-1~\mathrm{and}~\chi\left( \frac{2}{b}\right)=1,$$
which is equivalent to the condition
$$\chi\left( 4-2ab\right)=-1~\mathrm{and}~\chi\left( 2b\right)=1.$$
Similarly, for the reduced equation in Case III, when $u=1$,  only $x=-\frac{2}{b}$ can be a desired solution, and it is a desired solution iff
$$\chi\left( 4-2ab\right)=-1~\mathrm{and}~\chi\left( 2b\right)=-1.$$

When $u=-1$, the results can be derived in the same way as before.
\hfill$\square$

With the above preparations, we are able to calculate the differential uniformity of $f_u(x)$ for each $u\in \gf_{p^n}$. As we mentioned before, we only need to  consider the case $\mathcal{U}_0\setminus\{0\}$ due to the known results.  We first present the results for $u=\pm1\in \mathcal{U}_0\setminus\{0\}$ in Proposition \ref{0.1-1} below.

\begin{proposition}\label{0.1-1} Let $f_u(x)$ be the generalized Ness-Helleseth function defined in (\ref{mainfun}). Then, $f_{1}(x)$ and  $f_{-1}(x)$ have the same differential uniformity and differential spectrum.  Their differential uniformity is $\frac{p^n+1}{4}$, and the differential spectrum is given by
\begin{equation*}
\begin{split}[\,&\omega_0=\frac{(p^n-1)(p^n+1-\Gamma_{p,n})}{8},~\omega_1=\frac{(p^n-1)(2p^n-2+\Gamma_{p,n})}{4},\\
&\omega_2=\frac{(p^n-1)(p^n+1-\Gamma_{p,n})}{8},~\omega_3=0,~\cdots,~\omega_{\frac{p^n-3}{4}}=0,~\omega_{\frac{p^n+1}{4}}=(p^n-1)\,],
\end{split}
\end{equation*}	
where $\Gamma_{p,n}=\sum\limits_{x\in \gf_{p^n}}\chi\left(x(x+1)(x+4)\right)$.
\end{proposition}
{\em Proof:} Note that
$$f_u(-x)=-(-ux^{d_1}+x^{d_2})=-f_{-u}(x)$$
since $d_1=\frac{p^n-1}{2}-1$ is even and $d_2=p^n-2$ is odd. Thus, $f_u(x)$ and $f_{-u}(x)$ have the same differential uniformity and differential spectrum. Hence, we only need to investigate the differential properties of $f_1(x)$.  According to  \eqref{N1} and \eqref{lem3fuNa0} in Lemma \ref{lemforNa0}, when $u=1$, for each pair $(a,b)\in \gf_{p^n}^*\times \gf_{p^n}$, we have
\begin{equation}\label{pro1-f1}N_{1,1}(a,b)=\left\{
\begin{array}{ll}
1, &~ \mathrm{if}~b=a^{-1}(1\pm\chi(a)),\\
0, &~ \mathrm{otherwise},
\end{array} \right.
\end{equation}
and
\begin{equation}\label{pro1-f2}
N_{1,2}(a,0)=\frac{p^n-3}{4}.
\end{equation}

Next we determine $N_{1,2}(a,b)$ for $(a,b)\in \gf_{p^n}^*\times \gf_{p^n}^*$ with the help of Table \ref{newtable1}. When $u=1$, by Lemmas \ref{I-IV}, \ref{II-III} and \ref{II-IV}, we have the following results:

 (a) Case I in Table \ref{newtable1} has a desired solution if
\begin{equation}\label{pro1-caseI-solution}
\chi(a^2b^2-8ab)=1\,\,\mbox{\rm{and}}\,\, \chi(\frac{2a}{b})=-1
\end{equation}
and no desired solution otherwise;

(b) Case IV in Table \ref{newtable1} has no desired solution;

(c) Cases II in Table \ref{newtable1} contributes exactly one desired solution if
\begin{equation}\label{condi-for-u=1case2}
\chi\left(4-2ab\right)=-1~ \mathrm{and}~ \chi\left(2b\right)=1
\end{equation}
and no desired solution otherwise;

(d) Cases III contributes exactly one desired solution if
\begin{equation}\label{condi-for-u=1case3}
\chi\left(4-2ab\right)=-1~ \mathrm{and}~ \chi\left(2b\right)=-1
\end{equation}
and no desired solution otherwise.

\noindent Note that the conditions in \eqref{condi-for-u=1case2} and \eqref{condi-for-u=1case3} cannot hold simultaneously. Thus, we conclude that  for each $(a,b)\in \gf_{p^n}^*\times \gf_{p^n}^*$,
\begin{equation}\label{pro1-f3}
N_{1,2}(a,b)=\left\{\begin{array}{ll}
2, &~ \mathrm{if}~(a,b)~\mathrm{satisfies~\eqref{pro1-caseI-solution}~and~ \eqref{condi-for-u=1case2}, ~or ~ \eqref{pro1-caseI-solution}~and~ \eqref{condi-for-u=1case3}},\\
1, &~ \mathrm{if}~(a,b)~\mathrm{satisfies~only~one~of~the~conditions~}\\
&~\mathrm{in~\eqref{pro1-caseI-solution},~\eqref{condi-for-u=1case2} ~and~\eqref{condi-for-u=1case3}},\\
0, &~\mathrm{otherwise}.
\end{array} \right.
\end{equation}

By \eqref{N}, \eqref{pro1-f1}, \eqref{pro1-f2} and \eqref{pro1-f3}, we can get the values of $N_{1}(a,b)$  for each$(a,b)\in\gf_{p^n}^*\times\gf_{p^n}$ as follows
\begin{equation}\label{pro1-N}N_{1}(a,b)=\left\{
\begin{array}{ll}
\frac{p^n+1}{4}, &~\mathrm{if}~b=0,\\
2, &~ \mathrm{if}~(a,b)~\mathrm{satisfies~\eqref{pro1-caseI-solution}~and~ \eqref{condi-for-u=1case2}, ~or ~ \eqref{pro1-caseI-solution}~and~ \eqref{condi-for-u=1case3}},\\
1, &~\mathrm{if}~b=2a^{-1},\\
1, &~ \mathrm{if}~(a,b)~\mathrm{satisfies~only~one~of~the~conditions~}\\
&~\mathrm{in~\eqref{pro1-caseI-solution},~\eqref{condi-for-u=1case2} ~and~\eqref{condi-for-u=1case3}},\\
0, &~\mathrm{otherwise}.
\end{array} \right.
\end{equation}
Therefore, the differential uniformity of $f_1(x)$ is $\frac{p^n+1}{4}$ and the number of pairs $(a,b)\in\gf_{p^n}^*\times\gf_{p^n}$ such that $N_{1}(a,b)=\frac{p^n+1}{4}$ is equal to $p^n-1$.

In what follows, we determine the number of pairs $(a,b)\in \gf_{p^n}^*\times \gf_{p^n}$ such that  $N_1(a,b)=2$. According to \eqref{pro1-N}, $N_1(a,b)=2$ iff one of the following two conditions holds:

\noindent(i) $\chi(a^2b^2-8ab)=1$, $\chi(2ab)=-1$, $\chi(4-2ab)=-1$ and $\chi(2b)=1$,

\noindent(ii) $\chi(a^2b^2-8ab)=1$, $\chi(2ab)=-1$, $\chi(4-2ab)=-1$ and $\chi(2b)=-1$.

\noindent Note that one of them  holds iff
$$\chi(a^2b^2-8ab)=1, \chi(2ab)=-1~\mathrm{and}~\chi(2ab-4)=1.$$
Let $z=ab$, and let $M_1$ denote the number of elements $z \in\gf_{p^n}$ satisfying  the conditions $\chi(z^2-8z)=1$, $\chi(2z)=-1$ and $\chi(2z-4)=1$. Then we have
\begin{equation*}
\begin{split}
8M_1
=&\sum_{\substack{z\neq0,2,8}}\left(1+\chi(z^2-8z)\right)\left(1-\chi(2z)\right)\left(1+\chi(2z-4)\right)\\
=&\sum_{z\in\mathbb{F}_{p^n}}\left(1+\chi(z^2-8z)\right)\left(1-\chi(2z)\right)\left(1+\chi(2z-4)\right).
\end{split}
\end{equation*}

By Lemma \ref{charactersumquadratic}, we further get
\begin{equation*}
\begin{split}
8M_1=&p^n+1+\sum_{z\in\mathbb{F}_{p^n}}\chi((2z-4)(z^2-8z)).
\end{split}
\end{equation*}
Notice that \begin{equation*}
\begin{split}
&\sum_{z\in\mathbb{F}_{p^n}}\chi((2z-4)(z^2-8z))\\
=&\sum_{x\in\mathbb{F}_{p^n}}\chi((2(-2x)-4)(-2x)(-2x-8))\\
=&-\sum_{x\in\mathbb{F}_{p^n}}\chi(x(x+1)(x+4))\\
=&-\Gamma_{p,n}.
\end{split}
\end{equation*}
Thus, $8M_1=p^n+1-\Gamma_{p,n}$. The number of pairs $(a,b)\in \gf_{p^n}^*\times\gf_{p^n}$ such that $N_{1}(a,b)=2$ is equal to $(p^n-1)M_1=\frac{(p^n-1)(p^n+1-\Gamma_{p,n})}{8}$.
 Let $[\omega_0,\omega_1,\cdots,\omega_{\frac{p^n+1}{4}}]$ be the differential spectrum of $f_1(x)$. According the above arguments, we have known that
 $\omega_2=\frac{(p^n-1)(p^n+1-\Gamma_{p,n})}{8}$, $\omega_3=\cdots=\omega_{\frac{p^n-3}{4}}=0$, and $\omega_{\frac{p^n+1}{4}}=p^n-1$. Utilizing the identities \eqref{prop} yields the desire result.
\hfill$\square$

Next we compute the differential uniformity of $f_u(x)$ for $u\in\mathcal{U}_0\setminus\{0,\pm1\}$. The following lemma is needed.

\begin{lemma}\label{equicodi1}Let $u\in\mathcal{U}_0\setminus\{0,\pm1\}$ and $(a,b)\in \gf_{p^n}^*\times\gf_{p^n}$ satisfying $\chi(a^2b^2-4ab+4u^2)=1$. Then,
$\chi\left(2u^2-ab\pm u\sqrt{a^2b^2-4ab+4u^2}\right)=-1$ iff $\chi(2ab(1+\sqrt{1-u^2})-4u^2)=1$.
\end{lemma}
{\em Proof:} Note that $\chi(2u^2-ab\pm u\sqrt{a^2b^2-4ab+4u^2})=-1$ means that there exists  an element $y\in \mathbb{F}_{p^n}^*$ such that
\begin{equation*}\label{deg4eq0}
	2u^2-ab\pm u\sqrt{a^2b^2-4ab+4u^2}=-y^2.
\end{equation*}The above equation   is equivalent to
\begin{equation}\label{deg4eq}
	y^4+(4u^2-2ab)y^2-(u^2-1)a^2b^2=0.
\end{equation}
Since the discriminant $(4u^2-2ab)^2-4(-(u^2-1)a^2b^2)=4u^2(a^2b^2-4ab+4u^2)$ is a square in $\mathbb{F}_{p^n}^*$, \eqref{deg4eq}  can be rewritten as
\begin{equation*}
	(y^2-z_0)(y^2-z_1)=0,
\end{equation*}
where $z_0$ and $z_1$ are the two distinct roots of the quadratic equation $z^2+(4u^2-2ab)z-(u^2-1)a^2b^2=0$ in variable $z$. Note that $z_0z_1=-(u^2-1)a^2b^2$.  When $u\in\mathcal{U}_0\setminus \{0,\pm1\}$, we know that $\chi(u+1)=\chi(1-u)$ and thus $-(u^2-1)a^2b^2$ is square.  So  both $z_0$ and $z_1$ are square or neither of them is square.  Then, \eqref{deg4eq}  either has four distinct solutions in $\mathbb{F}_{p^n}$ or has no solution.  In particular, \eqref{deg4eq} has four distinct solutions if  and only if there exist two elements $c$ and $d$ of $\mathbb{F}_{p^n}$ such that
\begin{equation}\label{condiforfactor1}
\begin{split}
	&y^4+(4u^2-2ab)y^2-(u^2-1)a^2b^2\\
=&(y^2+cy+d)(y^2-cy+d)
\end{split}
\end{equation} with
\begin{equation}\label{condiforfactor}
	\left \{\begin{array}{lll}
		d^2=(1-u^2)a^2b^2,\\
		2d-c^2=4u^2-2ab,\\
		\chi(c^2-4d)=1.
	\end{array}\right.\end{equation}
Since $(1-u^2)a^2b^2$ is square,  from  \eqref{condiforfactor},  we  have
$$c^2=-4u^2+2ab+2d=-4u^2+2ab\pm 2ab\sqrt{1-u^2}$$ and
 $$c^2-4d=-4u^2+2ab-2d=-4u^2+2ab\mp 2ab\sqrt{1-u^2}.$$
Thus,  there exist $c$ and $d$ such that \eqref{condiforfactor1} and \eqref{condiforfactor} hold iff  $-4u^2+2ab\pm 2ab\sqrt{1-u^2}$ are square.  Since  $(-4u^2+2ab+2ab\sqrt{1-u^2})(-4u^2+2ab-2ab\sqrt{1-u^2})=4u^2(a^2b^2-4ab+4u^2)$, which is square, it follows that $-4u^2+2ab\pm 2ab\sqrt{1-u^2}$ are square iff one of them is square.  Then, we can conclude that \eqref{deg4eq} has four distinct solutions if and only if
$$\chi\left(-4u^2+2ab+2ab\sqrt{1-u^2}\right)=1.$$
Therefore,  $\chi\left(2u^2-ab\pm u\sqrt{a^2b^2-4ab+4u^2}\right)=-1$ iff $\chi\left(-4u^2+2ab+2ab\sqrt{1-u^2}\right)=1$.
\hfill$\square$

\begin{proposition}\label{4-uniform-con-1}
When $u\in\mathcal{U}_0\setminus\{0,\pm1\}$, the differential uniformity of the function $f_u(x)$ defined in (\ref{mainfun}), is at most $4$. Moreover, 
for the derivative equation $\mathbb{D}_af_u(x)=b$ of $f_u(x)$, one has the following results:

\noindent(i) it has four solutions iff there exists a pair  $(a,b)\in\gf_{p^n}^*\times\gf_{p^n}$ satisfying the following conditions
\begin{equation}\label{keycondi}
	\left \{\begin{array}{lll}
		\chi\left(\frac{a(u+1)}{b}\right)=-1,\\
		\chi\left(a^2b^2-4(u+1)ab\right)=1,\\
		\chi\left(a^2b^2+4(u-1)ab\right)=1,\\
		\chi\left(4u^2+a^2b^2-4ab\right)=1,\\
		\chi\left(-4u^2+2ab+2ab\sqrt{1-u^2}\right)=1;
\end{array}\right.\end{equation}
\noindent(ii) it has three solutions iff $(a,b)\in\gf_{p^n}^*\times\gf_{p^n}$ satisfies
\begin{equation*}\label{keycondiadd1}
	\left \{\begin{array}{lll}
		b=a^{-1}(1\pm u),\\
		\chi\left(4u^2+a^2b^2-4ab\right)=1,\\
		\chi\left(-4u^2+2ab+2ab\sqrt{1-u^2}\right)=1,
\end{array}\right.\end{equation*}
or
\begin{equation*}\label{keycondiadd2}
	\left \{\begin{array}{lll}
		\chi\left(\frac{a(u+1)}{b}\right)=-1,\\
		\chi\left(a^2b^2-4(u+1)ab\right)=1,\\
		\chi\left(a^2b^2+4(u-1)ab\right)\neq 1,\\
		\chi\left(4u^2+a^2b^2-4ab\right)=1,\\
		\chi\left(-4u^2+2ab+2ab\sqrt{1-u^2}\right)=1,
\end{array}\right.\end{equation*}
or
\begin{equation*}\label{keycondiadd3}
	\left \{\begin{array}{lll}
		\chi\left(\frac{a(u+1)}{b}\right)=-1,\\
		\chi\left(a^2b^2-4(u+1)ab\right)\neq 1,\\
		\chi\left(a^2b^2+4(u-1)ab\right)=1,\\
		\chi\left(4u^2+a^2b^2-4ab\right)=1,\\
		\chi\left(-4u^2+2ab+2ab\sqrt{1-u^2}\right)=1.
\end{array}\right.\end{equation*}

\end{proposition}
{\em Proof:} When $u\in\mathcal{U}_0\setminus\{0,\pm1\}$, for each pair $(a,b)\in\gf_{p^n}^*\times \gf_{p^n}$, by \eqref{N1} and \eqref{lem3fuNa0}, we have
\begin{equation}\label{pro2-N1}N_{u,1}(a,b)=\left\{
\begin{array}{ll}
1, &~ \mathrm{if}~b=a^{-1}(1\pm u\chi(a)),\\
0, &~ \mathrm{otherwise},
\end{array} \right.
\end{equation}
and
\begin{equation}\label{pro2-N2}
N_{u,2}(a,0)=0.
\end{equation}

Next we determine the values of $N_{u,2}(a,b)$ for $(a,b)\in \gf_{p^n}^*\times\gf_{p^n}^*$. When $u\in\mathcal{U}_0\setminus\{0,\pm1\}$, we have $\chi(u+1)=\chi(1-u)$ and $\chi\left(\frac{a(u+1)}{b}\right)=-\chi\left(\frac{a(u-1)}{b}\right)$.  It follows from  Lemma \ref{I-IV} that Cases I and IV in Table \ref{newtable1} contribute two desired solutions in total if
\begin{equation}\label{condiC1-4with4roots}
	\chi\left(\frac{a(u+1)}{b}\right)=-1,~\chi\left(a^2b^2-4(u+1)ab\right)=1 ~ \mathrm{and}~\chi\left(a^2b^2+4(u-1)ab\right)=1,
\end{equation}
one desired solution if
\begin{equation}\label{condiC1-4with4roots-ad1}
\begin{split}
	&\chi\left(\frac{a(u+1)}{b}\right)=-1,~\chi\left(a^2b^2-4(u+1)ab\right)=1 ~\mathrm{and}~\chi\left(a^2b^2+4(u-1)ab\right)\neq 1,\\
\mathrm{or}~&\chi\left(\frac{a(u+1)}{b}\right)=-1,~\chi\left(a^2b^2-4(u+1)ab\right)\neq 1~\mathrm{and}~\chi\left(a^2b^2+4(u-1)ab\right)= 1,\\
\end{split}
\end{equation}
and no desired solution otherwise. Note that if \eqref{condiC1-4with4roots} or \eqref{condiC1-4with4roots-ad1} holds, the condition \eqref{condiforCase2-3onlyonesolution} in Lemma \ref{II-IV} cannot hold.
By \eqref{condiC1-4with4roots}, \eqref{condiC1-4with4roots-ad1} and Lemma \ref{II-IV}, we can get the following conclusion:
when $u\in\mathcal{U}_0\setminus\{0,\pm1\}$ and $(a,b)\in\gf_{p^n}^*\times \gf_{p^n}^*$,
\begin{equation}\label{pro2-N2abn0}N_{u,2}(a,b)=\left\{
\begin{array}{ll}
4, &~ \mathrm{if~\eqref{condiC1-4with4roots}~and~\eqref{condiforCase2-3both-have-desired-solution}~in~Lemma~\ref{II-IV}~hold~simultaneously},\\
3, &~ \mathrm{if~\eqref{condiC1-4with4roots-ad1}~and~\eqref{condiforCase2-3both-have-desired-solution}~hold~simultaneously}\\
2, &~ \mathrm{\eqref{condiC1-4with4roots}~holds~and~\eqref{condiforCase2-3both-have-desired-solution}~does~not~hold},\\
&~\mathrm{or~\eqref{condiforCase2-3both-have-desired-solution}~holds~and~neither~\eqref{condiC1-4with4roots}~nor~\eqref{condiC1-4with4roots-ad1}~holds},\\
1, &~ \mathrm{if~\eqref{condiC1-4with4roots-ad1}~holds~and~\eqref{condiforCase2-3both-have-desired-solution}~does~not~hold,~or~\eqref{condiforCase2-3onlyonesolution}~holds },\\
0, &~ \mathrm{otherwise}.
\end{array} \right.
\end{equation}

Note that the condition $b=a^{-1}(1\pm u\chi(a))$ exactly means that $b=a^{-1}(1\pm u)$. When $u\in\mathcal{U}_0\setminus\{0,\pm1\}$, if $b=a^{-1}(1\pm u)$, then  $b\neq 0$ and $\chi\left(\frac{a(u+1)}{b}\right)=-\chi\left(\frac{a(u-1)}{b}\right)=1$. This implies  that if $b=a^{-1}(1\pm u)$, then neither \eqref{condiC1-4with4roots} nor \eqref{condiC1-4with4roots-ad1} holds. Thus, when $u\in\mathcal{U}_0\setminus\{0,\pm1\}$, by \eqref{pro2-N1}, \eqref{pro2-N2} and \eqref{pro2-N2abn0},  we have

\begin{equation}\label{pro2-Nuabgeneral}N_{u}(a,b)=\left\{
\begin{array}{ll}
4, &~ \mathrm{if~\eqref{condiC1-4with4roots}~and~\eqref{condiforCase2-3both-have-desired-solution}~hold~simultaneously},\\
3, &~ \mathrm{if}~b=a^{-1}(1\pm u)~\mathrm{and~\eqref{condiforCase2-3both-have-desired-solution}~holds},\\
&~\mathrm{or}~\mathrm{\eqref{condiC1-4with4roots-ad1}~and~\eqref{condiforCase2-3both-have-desired-solution}~hold~simultaneously},\\
2, &~ \mathrm{if}~b=a^{-1}(1\pm u)~\mathrm{and~\eqref{condiforCase2-3onlyonesolution}~holds},\\
&~\mathrm{or~}b\ne a^{-1}(1\pm u)~\mathrm{and~} N_{u,2}(a,b)=2,\\
1, &~ \mathrm{if}~b=a^{-1}(1\pm u)~\mathrm{and~neither~\eqref{condiforCase2-3both-have-desired-solution}~nor~\eqref{condiforCase2-3onlyonesolution}~holds},\\
&~\mathrm{or}~b\ne a^{-1}(1\pm u)~\mathrm{and~}N_{u,2}(a,b)=1,\\
0,&~\mathrm{otherwise},
\end{array} \right.
\end{equation}
where $(a,b)\in \gf_{p^n}^*\times\gf_{p^n}^*$ and $N_{u,2}(a,b)$ is given in \eqref{pro2-N2abn0}.

 The above arguments show that when $u\in\mathcal{U}_0\setminus\{0,\pm1\}$, we have $N_u(a,b)\in\{0,1,\cdots,4\}$. The conditions on $(a,b)$ such that  $N_u(a,b)=i$, $i=0,1,\cdots,4$, are given in  \eqref{pro2-Nuabgeneral}. This fact together with Lemma \ref{equicodi1} yields the desired result.
\hfill$\square$

When $u\in\mathcal{U}_0\setminus\{0,\pm1\}$, to check whether the differential uniformity of $f_u(x)$ is equal to $4$, we should verify that whether there exists a pair $(a,b)\in \gf_{p^n}^*\times \gf_{p^n}$ satisfying the conditions \eqref{keycondi}. Let $z=ab$, $\varphi(u)=2+2\sqrt{1-u^2}$, and $M$
denote the number of $z\in\gf_{p^n}$ satisfying the following conditions:
\begin{equation*}\label{keycondi0}
	\left \{\begin{array}{lll}
		\chi(-(u+1)z)=1,\\
		\chi\left(z^2-4(u+1)z\right)=1,\\
		\chi\left(z^2+4(u-1)z\right)=1,\\
		\chi(z^2-4z+4u^2)=1,\\
		\chi(\varphi(u)z-4u^2)=1.
	\end{array}\right.\end{equation*}
If we can show $M>0$, then there will exist $(p^n-1)M$ pairs $(a,b)\in\mathbb{F}_{p^n}^*\times \mathbb{F}_{p^n}$ such that the derivative equation $\mathbb{D}_af_u(x)=b$ has four solutions. We thereby prove that the differential uniformity of $f_u(x)$ is equal to $4$. For convenience, set
\begin{equation}\label{rep of fivep}
	\left \{\begin{array}{lll}
		g_1(z)=-(u+1)z,\\
		g_2(z)=z^2-4(u+1)z,\\
        g_3(z)=z^2+4(u-1)z,\\
        g_4(z)=z^2-4z+4u^2,\\
        g_5(z)=\varphi(u)z-4u^2.\\
\end{array}\right.\end{equation}
The roots of these fives polynomials are given in the following set
$$ \mathcal{R}=\left\{0, 4(1\pm u), 2\pm 2\sqrt{1-u^2}\right\}.$$
When  $u\in\mathcal{U}_0\setminus\{0,\pm1\}$,  the elements $0$, $4(1\pm u)$ are pairwise distinct, and the elements $0$, $2\pm 2\sqrt{1-u^2}$  are also pairwise distinct. However, when $u=\pm \frac{4}{5}$ (they belong to $\mathcal{U}_0$), $\{2\pm 2\sqrt{1-u^2}\}\cap \{4(1\pm u)\}=\{\frac{4}{5}\}$.  This can happen only if  $p>3$ since $u=\pm \frac{4}{5}=\mp 1$ when $p=3$. Thus, if $u\in\mathcal{U}_0\setminus\{0,\pm1,\pm\frac{4}{5}\}$, the elements in $\mathcal{R}$ are pairwise distinct. When calculating the value of $M$, it is beneficial for us to know the relationships of equality among the elements of $\mathcal{R}$. Thus, in the following we
first determine the differential uniformity of $f_u(x)$ for $u\in\mathcal{U}_0\setminus\{0,\pm1,\pm\frac{4}{5}\}$.

To clearly present our results, we introduce the set  $\mathcal{A}$ containing the $3$-tuples $(p,n,u)$ given in Table \ref{table3}, where the elements $u$ belong to $\mathcal{U}_0\setminus\{0,\pm1,\pm\frac{4}{5}\}$ for the given $(p,n)$.
\begin{table}[h]
\caption{The $3$-tuple $(p,n,u)$ set $\mathcal{A}$}\label{table3}
\centering
\setlength{\tabcolsep}{3mm}
\scriptsize
\begin{tabular}{|c|c|c|}
		\hline
		$(p,n,u)$ &  $(p,n,u)$ & $(p,n,u)$ \\  \hline
		 $(11,1,\pm5)$ & $(19,1,\pm2)$  & $(23,1,\pm4) $ \\  \hline
		 $ (31,1,\pm10) $&$(31,1,\pm13)$ & $ (47,1,\pm11) $\\ \hline
		  $ (59,1,\pm15)$  &$ (71,1,\pm13) $& $(83,1,\pm4) $\\ \hline
         $(83,1,\pm38)$ &$  (151,1,\pm22) $ & -\\ \hline
\end{tabular}
\end{table}

\begin{proposition}\label{4-uniform} Let $\mathcal{A}$ be the set of the $3$-tuples $(p,n,u)$ listed in Table \ref{table3}, and $f_u(x)$ be the generalized Ness-Helleseth function defined in (\ref{mainfun}).
When $u\in\mathcal{U}_0\setminus\{0,\pm1,\pm\frac{4}{5}\}$, the differential uniformity of $f_u(x)$ is given as follows
 \begin{equation*}\label{dfoffu3}
 	\delta(f_u)=\left\{
 	\begin{array}{ll}
 		3, &~ \mathrm{if}~(p,n,u) \in \mathcal{A},\\
 		4, &~ \mathrm{otherwise}.
 	\end{array} \right.
 \end{equation*}

\end{proposition}
{\em Proof:}  We can compute $M$ by the following character sum
\begin{equation*}\label{compforN1}
32M=\sum_{z\in \mathbb{F}_{p^n} \setminus \mathcal{R}}(1+\chi(g_1))\times(1+\chi(g_2))\times(1+\chi(g_3))
	\times(1+\chi(g_4))\times(1+\chi(g_5)).
\end{equation*}
Note that $1+\chi(g_5(0))=0$ and $1+\chi(g_1(4(1\pm u)))=0$. In addition, we can prove that $\chi(\varphi(u)(u+1))=1$ in the same way
as Lemma \ref{equicodi1}. Thus, when $z=2+2\sqrt{1-u^2}=\varphi(u)$, we have $1+\chi(g_1(\varphi(u)))=0$. For $z=2-2\sqrt{1-u^2}=\frac{4u^2}{\varphi(u)}$, we have $1+\chi\left(g_1(2-2\sqrt{1-u^2})\right)=1+\chi\left(g_1(-\frac{4u^2(u+1)}{\varphi(u)})\right)=0$. Thus, we have
\begin{equation*}
\sum_{z\in \mathcal{R}}(1+\chi(g_1))\times(1+\chi(g_2))\times(1+\chi(g_3))
	\times(1+\chi(g_4))\times(1+\chi(g_5))=0
\end{equation*}
and
\begin{equation}\label{vaule-M}32M=\sum_{z\in \mathbb{F}_{p^n}}(1+\chi(g_1))\times(1+\chi(g_2))\times(1+\chi(g_3))\times(1+\chi(g_4))\times(1+\chi(g_5)).\end{equation}

Expanding \eqref{vaule-M} yields $32$ terms, and each term is of the form $\sum\limits_{z\in \mathbb{F}_{p^n}}\chi\left(\prod\limits_{i \in I}g_i\right)$, where $I$ is an arbitrary subset of $\{1,2,3,4,5\}$. If $I$ is an empty set, we have $\sum\limits_{z\in \mathbb{F}_{p^n}}\chi\left(\prod\limits_{i \in I}g_i\right)=\sum\limits_{z\in \mathbb{F}_{p^n}}\chi(1)=p^n$. We list all these terms  in Table \ref{table-6}.
According to \eqref{rep of fivep}, one can see that $g_1(z)$ and $g_5(z)$ are linear polynomials, while $g_2(z)$, $g_3(z)$ and $g_4(z)$ are quadratic polynomials, all of which can be factorized into a product of two linear polynomials.  In the canonical factorizations of $g_i(z)$ ($i=1,2,\cdots,5$), there are five linear factors $z$, $ z-4(1\pm u)$, $z\pm(2+2\sqrt{1-u^2})$, and the linear factors $z-(2+2\sqrt{1-u^2})$, $ z-4(u+1)$, $z+4(u-1)$ only
occur once. The five linear factors are pairwise distinct when $u\in\mathcal{U}_0\setminus\{0,\pm1,\pm\frac{4}{5}\}$. With these information, we can use Theorem \ref{weilbound} to estimate the character sum
$\sum\limits_{z\in \mathbb{F}_{p^n}}\chi\left(\prod\limits_{i \in I}g_i\right)$.  The details of the computations are listed in Tables \ref{table-6} and \ref{table-7}. Here we compute the character sum $\sum\limits_{z \in \gf_{p^n}}\chi(g_1g_3g_4)$ (No. 20 in Tables \ref{table-6} and \ref{table-7}) as an example. By Table \ref{table-6}, we have
\[
\begin{split}
\sum\limits_{z \in \gf_{p^n}}\chi(g_1g_3g_4)=&\sum\limits_{z \in \gf_{p^n}}\chi\left(-(u+1)(z+4(u-1))(z-2-2\sqrt{1-u^2})(z-2+2\sqrt{1-u^2})z^2\right)\\
=&\sum\limits_{z \in \gf_{p^n}^*}\chi\left(-(u+1)(z+4(u-1))(z-2-2\sqrt{1-u^2})(z-2+2\sqrt{1-u^2})\right) \\
=&\sum\limits_{z \in \gf_{p^n}}\chi\left(-(u+1)(z+4(u-1))(z-2-2\sqrt{1-u^2})(z-2+2\sqrt{1-u^2}))\right)\\
&+\chi\left(16u^2(u^2-1)\right), \\
\end{split}
\]
Note that $\chi\left(16u^2(u^2-1)\right)=-1$ since $\chi(u+1)=\chi(1-u)$, and utilizing  Theorem \ref{weilbound}, we have
$$\bigg\lvert\sum\limits_{z \in \gf_{3^n}}\chi\left(-(u+1)(z+4(u-1))(z-2-2\sqrt{1-u^2})(z-2+2\sqrt{1-u^2})\right)\bigg\lvert \leq 2\sqrt{p^n}.$$

\begin{table}
\caption{Terms of the polynomial $\prod\limits_{i=1}^5(1+\chi(g_i(z)))$ }\label{table-6}
\centering
\setlength{\tabcolsep}{3mm}
\scriptsize
\begin{tabular}{|c|c|c|}
		\hline
		No. & Term &Canonical  Factorization  \\  \hline
        1  &   $1$       & $1$\\  \hline
		2  & $g_1$ & $-(u+1)z$\\  \hline
		3  & $g_2$ & $(z-4(u+1))z$\\  \hline
		4  & $g_3$ & $(z+4(u-1))z$\\  \hline
		5  & $g_4$ & $(z-2-2\sqrt{1-u^2})(z-2+2\sqrt{1-u^2})$\\  \hline
        6  & $g_5$ & $\varphi(u)(z-2+2\sqrt{1-u^2})$\\  \hline
        7  & $g_1g_2$ & $-(u+1)(z-4(u+1))z^2$\\  \hline
		8  & $g_1g_3$ & $-(u+1)(z+4(u-1))z^2$\\  \hline
		9  & $g_1g_4$ & $-(u+1)(z-2-2\sqrt{1-u^2})(z-2+2\sqrt{1-u^2})z$\\  \hline
		10 & $g_1g_5$ & $-\varphi(u)(u+1)(z-2+2\sqrt{1-u^2})z$\\  \hline
        11 & $g_2g_3$ & $(z-4(u+1))(z+4(u-1))z^2$\\  \hline
        12 & $g_2g_4$ & $(z-4(u+1))(z-2-2\sqrt{1-u^2})(z-2+2\sqrt{1-u^2})z$\\  \hline
        13 & $g_2g_5$ & $\varphi(u)(z-4(u+1))(z-2+2\sqrt{1-u^2})z$\\  \hline
        14 & $g_3g_4$ & $(z+4(u-1))(z-2-2\sqrt{1-u^2})(z-2+2\sqrt{1-u^2})z$\\  \hline
        15 & $g_3g_5$ & $\varphi(u)(z+4(u-1))(z-2+2\sqrt{1-u^2})z$\\  \hline
        16 & $g_4g_5$ & $\varphi(u)(z-2-2\sqrt{1-u^2})(z-2+2\sqrt{1-u^2})^2$\\  \hline
        17 & $g_1g_2g_3$ & $-(u+1)(z-4(u+1))(z+4(u-1))z^3$\\  \hline
		18 & $g_1g_2g_4$ & $-(u+1)(z-4(u+1))(z-2-2\sqrt{1-u^2})(z-2+2\sqrt{1-u^2})z^2$\\  \hline
		19 & $g_1g_2g_5$ & $-\varphi(u)(u+1)(z-4(u+1))(z-2+2\sqrt{1-u^2})z^2$\\  \hline
		20 & $g_1g_3g_4$ & $-(u+1)(z+4(u-1))(z-2-2\sqrt{1-u^2})(z-2+2\sqrt{1-u^2})z^2$\\  \hline
        21 & $g_1g_3g_5$ & $-\varphi(u)(u+1)(z+4(u-1))(z-2+2\sqrt{1-u^2})z^2$\\  \hline
        22 & $g_1g_4g_5$ & $-\varphi(u)(u+1)(z-2-2\sqrt{1-u^2})(z-2+2\sqrt{1-u^2})^2z$\\  \hline
        23 & $g_2g_3g_4$ & $(z-4(u+1))(z+4(u-1))(z-2-2\sqrt{1-u^2})(z-2+2\sqrt{1-u^2})z^2$\\  \hline
        24 & $g_2g_3g_5$ & $\varphi(u)(z-4(u+1))(z+4(u-1))(z-2+2\sqrt{1-u^2})z^2$\\  \hline
        25 & $g_2g_4g_5$ & $\varphi(u)(z-4(u+1))(z-2-2\sqrt{1-u^2})(z-2+2\sqrt{1-u^2})^2z$\\  \hline
        26 & $g_3g_4g_5$ & $\varphi(u)(z+4(u-1))(z-2-2\sqrt{1-u^2})(z-2+2\sqrt{1-u^2})^2z$\\  \hline
        27 & $g_1g_2g_3g_4$ & $-(u+1)(z-4(u+1))(z+4(u-1))(z-2-2\sqrt{1-u^2})(z-2+2\sqrt{1-u^2})z^3$\\  \hline
        28 & $g_1g_2g_3g_5$ & $-\varphi(u)(u+1)(z-4(u+1))(z+4(u-1))(z-2+2\sqrt{1-u^2})z^3$\\  \hline
        29 & $g_1g_2g_4g_5$ & $-\varphi(u)(u+1)(z-4(u+1))(z-2-2\sqrt{1-u^2})(z-2+2\sqrt{1-u^2})^2z^2$\\  \hline
        30 & $g_1g_3g_4g_5$ & $-\varphi(u)(u+1)(z+4(u-1))(z-2-2\sqrt{1-u^2})(z-2+2\sqrt{1-u^2})^2z^2$\\  \hline
        31 & $g_2g_3g_4g_5$ & $\varphi(u)(z-4(u+1))(z+4(u-1))(z-2-2\sqrt{1-u^2})(z-2+2\sqrt{1-u^2})^2z^2$\\  \hline
        32 & $g_1g_2g_3g_4g_5$ & $-\varphi(u)(u+1)(z-4(u+1))(z+4(u-1))(z-2-2\sqrt{1-u^2})(z-2+2\sqrt{1-u^2})^2z^3$\\  \hline
\end{tabular}
\end{table}

 \begin{table}
\caption{The values or bounds of the character sums}\label{table-7}
\centering
\setlength{\tabcolsep}{3mm}
\small
\begin{tabular}{|c|c|c|}
		\hline
		No.  & Character sum & Value ~or ~Bound\\  \hline
		1    & $\sum_{z \in \gf_{p^n}}\chi(1)$                 & $p^n$\\  \hline
        2    & $\sum_{z \in \gf_{p^n}}\chi(g_1)$               & $0$\\  \hline
        3    & $\sum_{z \in \gf_{p^n}}\chi(g_2)$               & $-1$\\  \hline
        4    & $\sum_{z \in \gf_{p^n}}\chi(g_3)$               & $-1$\\  \hline
        5    & $\sum_{z \in \gf_{p^n}}\chi(g_4)$               & $-1$\\  \hline
        6    & $\sum_{z \in \gf_{p^n}}\chi(g_5)$               & $0$\\  \hline
        7    & $\sum_{z \in \gf_{p^n}}\chi(g_1g_2)$            & $-1$\\  \hline
        8    & $\sum_{z \in \gf_{p^n}}\chi(g_1g_3)$            & $-1$\\  \hline
        9    & $\sum_{z \in \gf_{p^n}}\chi(g_1g_4)$            & $[-2\sqrt{p^n},2\sqrt{p^n}]$\\  \hline
        10   & $\sum_{z \in \gf_{p^n}}\chi(g_1g_5)$            & $1$\\  \hline
        11   & $\sum_{z \in \gf_{p^n}}\chi(g_2g_3)$            & $-2$\\  \hline
        12   & $\sum_{z \in \gf_{p^n}}\chi(g_2g_4)$            & $[-3\sqrt{p^n},3\sqrt{p^n}]$\\  \hline
        13   & $\sum_{z \in \gf_{p^n}}\chi(g_2g_5)$            & $[-2\sqrt{p^n},2\sqrt{p^n}]$\\  \hline
        14   & $\sum_{z \in \gf_{p^n}}\chi(g_3g_4)$            & $[-3\sqrt{p^n},3\sqrt{p^n}]$\\  \hline
        15   & $\sum_{z \in \gf_{p^n}}\chi(g_3g_5)$            & $[-2\sqrt{p^n},2\sqrt{p^n}]$\\  \hline
        16   & $\sum_{z \in \gf_{p^n}}\chi(g_4g_5)$            & $[-1,1]$\\  \hline
        17   & $\sum_{z \in \gf_{p^n}}\chi(g_1g_2g_3)$         & $[-2\sqrt{p^n},2\sqrt{p^n}]$\\  \hline
        18   & $\sum_{z \in \gf_{p^n}}\chi(g_1g_2g_4)$         & $[-2\sqrt{p^n}-1,2\sqrt{p^n}-1]$\\  \hline
        19   & $\sum_{z \in \gf_{p^n}}\chi(g_1g_2g_5)$         & $2$\\  \hline
        20   & $\sum_{z \in \gf_{p^n}}\chi(g_1g_3g_4)$         & $[-2\sqrt{p^n}-1,2\sqrt{p^n}-1]$\\  \hline
        21   & $\sum_{z \in \gf_{p^n}}\chi(g_1g_3g_5)$         & $2$\\  \hline
        22   & $\sum_{z \in \gf_{p^n}}\chi(g_1g_4g_5)$         & $[0,2]$\\  \hline
        23   & $\sum_{z \in \gf_{p^n}}\chi(g_2g_3g_4)$         & $[-3\sqrt{p^n}-1,3\sqrt{p^n}-1]$\\  \hline
        24   & $\sum_{z \in \gf_{p^n}}\chi(g_2g_3g_5)$         & $[-2\sqrt{p^n}+1,2\sqrt{p^n}+1]$\\  \hline
        25   & $\sum_{z \in \gf_{p^n}}\chi(g_2g_4g_5)$         & $[-2\sqrt{p^n}-1,2\sqrt{p^n}+1]$\\  \hline
        26   & $\sum_{z \in \gf_{p^n}}\chi(g_3g_4g_5)$         & $[-2\sqrt{p^n}-1,2\sqrt{p^n}+1]$\\  \hline
        27   & $\sum_{z \in \gf_{p^n}}\chi(g_1g_2g_3g_4)$      & $[-4\sqrt{p^n},4\sqrt{p^n}]$\\  \hline
        28   & $\sum_{z \in \gf_{p^n}}\chi(g_1g_2g_3g_5)$      & $[-3\sqrt{p^n},3\sqrt{p^n}]$\\  \hline
        29   & $\sum_{z \in \gf_{p^n}}\chi(g_1g_2g_4g_5)$      & $[1,3]$\\  \hline
        30   & $\sum_{z \in \gf_{p^n}}\chi(g_1g_3g_4g_5)$      & $[1,3]$\\  \hline
        31   & $\sum_{z \in \gf_{p^n}}\chi(g_2g_3g_4g_5)$      & $[-2\sqrt{p^n},2\sqrt{p^n}+2]$\\  \hline
        32   & $\sum_{z \in \gf_{p^n}}\chi(g_1g_2g_3g_4g_5)$   & $[-3\sqrt{p^n}-1,3\sqrt{p^n}+1]$\\  \hline
\end{tabular}
\end{table}

\noindent Thus, we get the bound $$-2\sqrt{p^n}-1\leq\sum\limits_{z \in \gf_{p^n}}\chi(g_1g_3g_4)\leq 2\sqrt{p^n}-1,$$
which is written as the  interval  $[-2\sqrt{p^n}-1, 2\sqrt{p^n}-1]$ in Table \ref{table-7}.  For any character sum $\sum\limits_{z\in \mathbb{F}_{p^n}}\chi\left(\prod\limits_{i \in I}g_i\right)$  in Table \ref{table-7}, if its exact value cannot be determined, a bound for it will be given in the form of an interval. When evaluating  the sums, we frequently use the fact that $\chi(\varphi(u)(u+1))=1$ if $u\in\mathcal{U}_0\setminus\{0,\pm1\}$.
Using the bounds and values in  Table \ref{table-7}, we can derive the following lower bound for \eqref{vaule-M}:
$$32M\geq p^n-6-39\sqrt{p^n}.$$
To make sure that $M>0$, one needs that $p^n>\left\lfloor\left( \frac{\sqrt{1545}+39}{2} \right)^2\right\rfloor=1533$. Thus, we have $\delta(f_u)=4$ when $p^n>1533$, $p^n\equiv3\pmod4$ and $u\in \mathcal{U}_0\setminus \{0,\pm1,\pm\frac{4}{5}\}$.

It remains to consider the case where $p^n\leq 1533$, $p^n\equiv3\pmod4$ and $u\in \mathcal{U}_0\setminus \{0,\pm1,\pm\frac{4}{5}\}$. There are finite primes $p$ and integers $n$ such that $p^n\leq 1533$ and $p^n\equiv3\pmod4$. For each given such $p$ and $n$, there are also finite elements in $\mathcal{U}_0\setminus \{0,\pm1,\pm\frac{4}{5}\}$. So we can use Magma to compute the differential uniformity of $f_u(x)$ in this case. By Magma, we get that $\delta(f_u)=3$ for the $3$-tuples $(p,n,u)$ given in Table \ref{table3} and  $\delta(f_u)=4$ otherwise. This completes the proof.
 \hfill$\square$

Now we consider the differential uniformity of $f_u(x)$ for $u\in\{\pm\frac{4}{5}\}$ when $p>3$.

\begin{proposition}\label{3-uniform-4/5}
When $p>3$ and $u=\pm\frac{4}{5}$, the generalized Ness-Helleseth function $f_u(x)$ defined in (\ref{mainfun}) is differentially $3$-uniform.
\end{proposition}
{\em Proof:} Since $f_u(x)$ and $f_{-u}(x)$ have the same differential properties, we only consider $u=\frac{4}{5}$. We still use Proposition \ref{4-uniform-con-1} to determine the exact differential uniformity of $f_{4/5}(x)$ since $u=\frac{4}{5}\in \mathcal{U}_0\setminus\{0,\pm1\}$.

When $u=\frac{4}{5}$, the conditions (\ref{keycondi}) in Proposition \ref{4-uniform-con-1} can be rewritten as
\begin{equation}\label{keycondi-4/5}
	\left \{\begin{array}{lll}
		\chi(\frac{9}{5}ab)=-1,\\
		\chi\left(a^2b^2-\frac{36}{5}ab\right)=1,\\
		\chi\left(a^2b^2-\frac{4}{5}ab\right)=1,\\
		\chi(a^2b^2-4ab+\frac{64}{25})=1,\\
		\chi(\frac{16}{5}ab-\frac{64}{25})=1.
\end{array}\right.\end{equation}
By the first equation and the last one of \eqref{keycondi-4/5}, we have
$$\chi\left(\frac{9}{5}ab\right)\cdot\chi\left(\frac{16}{5}ab-\frac{64}{25}\right)=\chi\left(a^2b^2-\frac{4}{5}ab\right)=-1,$$
which contradicts the third equation $\chi\left(a^2b^2-\frac{4}{5}ab\right)=1$. Thus, there is no pair $(a,b)\in \gf_{p^n}^*\times \gf_{p^n}$ satisfying \eqref{keycondi-4/5}, and the differential uniformity of $f_{4/5}(x)$  is at most $3$.

Next we check whether the differential uniformity of  $f_{4/5}(x)$ is equal to $3$. By Proposition \ref{4-uniform-con-1} (ii), this means that we should check whether there exists a pair $(a,b)\in \gf_{p^n}^*\times \gf_{p^n}$ such that\begin{equation}\label{keycondi-4/5a}
	\left \{\begin{array}{lll}
		ab=\frac{9}{5}~\mathrm{or}~\frac{1}{5},\\		
		\chi\left(a^2b^2-4ab+\frac{64}{25}\right)=1,\\
		\chi\left(\frac{16}{5}ab-\frac{64}{25}\right)=1,
\end{array}\right.
\end{equation}
or
\begin{equation}\label{keycondi-4/5c}
	\left \{\begin{array}{lll}
        \chi\left(\frac{9ab}{5}\right)=-1,\\
        \chi\left(a^2b^2-\frac{36}{5}ab\right)= 1,\\
        \chi\left(a^2b^2-\frac{4}{5}ab\right)\neq 1,\\
		\chi\left(a^2b^2-4ab+\frac{64}{25}\right)=1,\\
		\chi\left(\frac{16}{5}ab-\frac{64}{25}\right)=1,\\
\end{array}\right.\end{equation}
or
\begin{equation}\label{keycondi-4/5b}
	\left \{\begin{array}{lll}
        \chi\left(\frac{9ab}{5}\right)=-1,\\
        \chi\left(a^2b^2-\frac{36}{5}ab\right)\neq 1,\\
        \chi\left(a^2b^2-\frac{4}{5}ab\right)=1,\\
	    \chi\left(a^2b^2-4ab+\frac{64}{25}\right)=1,\\
		\chi\left(\frac{16}{5}ab-\frac{64}{25}\right)=1.\\
\end{array}\right.
\end{equation}
Note that \eqref{keycondi-4/5b} cannot hold due to the same reason as \eqref{keycondi-4/5}. Thus, we only need to check the conditions in \eqref{keycondi-4/5a} and \eqref{keycondi-4/5c}. If $ab=\frac{9}{5}$, \eqref{keycondi-4/5a} holds iff $\chi(5)=\chi(-7)=1$; if $ab=\frac{1}{5}$, \eqref{keycondi-4/5a} holds iff $\chi(5)=\chi(-3)=1$. So the conditions in \eqref{keycondi-4/5a} depend only
on the characteristic $p$. We turn to the conditions in \eqref{keycondi-4/5c}, which is equivalent to
  \begin{equation}\label{keycondi-4/5d}
	\left \{\begin{array}{lll}
        \chi\left(\frac{ab}{5}\right)=-1,\\
        \chi\left(a^2b^2-\frac{36}{5}ab\right)= 1,\\
		\chi\left(a^2b^2-4ab+\frac{64}{25}\right)=1,\\
		\chi\left(\frac{1}{5}ab-\frac{4}{25}\right)=1.
\end{array}\right.\end{equation}
 Let $ab=\frac{z}{5}$ and \eqref{keycondi-4/5d} can be rewritten as
 \begin{equation}\label{keycondi-4/5e}
	\left \{\begin{array}{lll}
        \chi(z)=-1,\\
        \chi\left(z^2-36z\right)= 1,\\
		\chi\left(z^2-20z+64\right)=1,\\
		\chi\left(z-4\right)=1.\\        
\end{array}\right.
\end{equation}
Let $N$ denote the number of $z\in\gf_{p^n}$ satisfying \eqref{keycondi-4/5e}. To show there exist pairs $(a,b)\in \gf_{p^n}^*\times \gf_{p^n}$ satisfying \eqref{keycondi-4/5d}, it suffices to prove that $N>0$.  Similarly, we have
\begin{equation*}
16N=\sum_{z\in \mathbb{F}_{p^n}}(1-\chi(h_1(z)))\times(1+\chi(h_2(z)))\times(1+\chi(h_3(z)))
	\times(1+\chi(h_4)),
\end{equation*}
where $h_1(z)=z$ and $h_2(z)=z^2-36z$, $h_3(z)=z^2-20z+64$ and $h_4(z)=z-4$. Using the same techniques as Proposition \ref{4-uniform}, we can show that
\begin{equation*}
16N\geq p^n-2-11\sqrt{p^n}.
\end{equation*}
Note that $N>0$ if $p^n>\left\lfloor\left(\frac{11+\sqrt{129}}{2}\right)^2 \right\rfloor=124$. Therefore, when $u=\frac{4}{5}$, the differential uniformity of $f_u(x)$ is $3$ if $p^n>124$ and $p^n\equiv3\pmod4$. For the remaining case where $p^n\leq 124$ and $p^n\equiv3\pmod4$, the differential uniformity of  $f_{4/5}(x)$ can be directly computed by Magma and it is always equal to  $3$.
\hfill$\square$

The following theorem is an immediate consequence of  Propositions \ref{0.1-1}, \ref{4-uniform} and \ref{3-uniform-4/5}.

\begin{theorem}\label{maintheorm}
Let $n$ be a positive integer and $p$ be an odd prime satisfying $p^n\equiv3\pmod4$  and $p^n\ge 7$, and $\mathcal{U}_0$ be the set defined in \eqref{setu}. When $u\in\mathcal{U}_0\setminus \{0\}$, the differential uniformity of the generalized Ness-Helleseth function $f_u(x)$ defined  in (\ref{mainfun}) is given as follows
\begin{equation*}\delta(f_u)=\left\{
\begin{array}{ll}
\frac{p^n+1}{4}, &~ \mathrm{if}~u=\pm1,\\
3, &~ \mathrm{if}~p>3~\mathrm{and}~u=\pm\frac{4}{5},\\
3, &~ \mathrm{if}~(p,n,u)\in \mathcal{A} ,\\
4, &~ \mathrm{otherwise},\\
\end{array} \right.
\end{equation*}
where $\mathcal{A}$ is the set of the $3$-tuples $(p,n,u)$ given in Table \ref{table3}.
\end{theorem}

According to Theorems \ref{result of zeng}, \ref{result of zha}, \ref{resultofu=0} and \ref{maintheorm}, we have following corollary.
\begin{corollary}\label{c-1}
Let $n$ be an odd integer and $p$ be an odd prime with $p^n\equiv3\pmod4$  and $p^n\ge 7$.  Then, the generalized Ness-Helleseth function  $f_u(x)$ defined in (\ref{mainfun}) is an APN function iff $n$, $p$ and $u\in\gf_{p^n}$ satisfy one of the following conditions:

\noindent(i) $p^n \equiv2 \pmod3$ and $u=0$;

\noindent(ii)  $u\in\mathcal{U}_{10}\cup \mathcal{U}_{11}$, where $\mathcal{U}_{10}$ and $\mathcal{U}_{11}$ are defined in \eqref{setu};

\noindent(iii) $p=7$, $n=1$ and $u=\pm1$.
\end{corollary}

\section{The differential spectrum of $f_u(x)$ when $\chi(u+1)=\chi(u-1)$}\label{main-resultforFp-2}
Recall the definition of the sets $\mathcal{U}_i$ and $\mathcal{U}_{1j}$ from \eqref{setu}, where $i=0,1$ and $j=0,1,2$. Note that $\mathcal{U}_1=\left(\mathcal{U}_{10}\cup \mathcal{U}_{11}\right)\cup \mathcal{U}_{12}$ and $\left(\mathcal{U}_{10}\cup \mathcal{U}_{11}\right)\cap \mathcal{U}_{12}=\emptyset$. In this section, we aim to determine the differential spectrum of $f_u(x)$ for $u\in\mathcal{U}_1$, i.e., $\chi(u+1)=\chi(u-1)$. To this end, we need to make some preparations. By Lemma \ref{lem cyclotomic}, we have $|\mathcal{U}_1|=\frac{p^n-3}{2}$, and by Lemma \ref{charactersumquadratic},  $|\mathcal{U}_{10}\cup \mathcal{U}_{11} |=\frac{p^n-1+2\chi(5)}{4}$. For $u\in \mathcal{U}_1$, let $T_1(u)$  denote the number  of $z\in \mathbb{F}_{p^n}^*$ satisfying
\begin{equation}\label{cond1}
	\chi((u+1)z)=-1,~\chi(z^2-4(u+1)z)=1~\mbox{ and}~\chi(z^2-4z+4u^2)=1
\end{equation}
and  $T_2(u)$  denote the number of $z\in \mathbb{F}_{p^n}^*$ satisfying   
\begin{equation}\label{cond2}
	\chi((u+1)z)=1,~\chi(z^2+4(u-1)z)=1~\mbox{ and}~\chi(z^2-4z+4u^2)=1.
\end{equation}
Let $\Gamma_0(u)$, $\Gamma_1(u)$ and $\Gamma_2(u)$ be the quadratic character sums defined by
\begin{equation}\label{defofGm}
		\left \{\begin{array}{lll}    
    \Gamma_0(u)&=&\sum\limits_{x\in \gf_{p^n}}\chi\left((u+1)x^3-4(u+1)x^2+4u^2(u+1)\right),\\
	\Gamma_1(u)&=&\sum\limits_{x\in \gf_{p^n}}\chi\left((u+1)^2x^3+(u^2-2u-2)x^2+(1-u^2)x\right),\\
	\Gamma_2(u)&=&\sum\limits_{x \in \gf_{p^n}}\chi\left((u+1)x^3-4(u+2)(u+1)x^2\right.\\
&&~~~~~~~~~~~\left.+4(u+2)^2(u+1)x-16u^2(u+1)^2\right).
\end{array}\right.
\end{equation}
The following proposition is a generalization of Proposition 4 in \cite{Xia-Bao2023}, and it can be proved in a similar manner.

\begin{proposition}\label{number}
With the notation as above, let $u\in \mathcal{U}_1$, then we have
$$T_1(-u)=T_2(u)$$
and
\[
\begin{split}
8T_1(u)&=p^n-7-\Gamma_0(u)+\Gamma_1(u)-\Gamma_2(u),\end{split}
\]
where $\Gamma_i(u)$, $i=0,1,2$, are defined as in \eqref{defofGm}.
\end{proposition}
{\em Proof:} Note that if $u$ satisfies $\chi(u+1)=\chi(u-1)$, so does $-u$. Replacing  $u$ by $-u$ in (\ref{cond1}), we can get the conditions (\ref{cond2}). Thus, we get
$T_1(-u)=T_2(u)$. So we only deal with $T_1(u)$. Let $\Omega=\{0,4(u+1)\}$ that contains  all the roots of $(u+1)z$ and $z^2-4(u+1)z$. Note that $z^2-4z+4u^2$ is irreducible over $\gf_{p^n}$ when $\chi(u+1)=\chi(u-1)$. Then, we have
\[
\begin{split}
8T_1(u) =&\sum_{z \in \gf_{p^n}\setminus\Omega} \left(1-\chi((u+1)z)\right)\left(1+\chi(z^2-4(u+1)z)\right)\left(1+\chi(z^2-4z+4u^2)\right)\\
     =&-2+\sum_{z \in \gf_{p^n}}\left(1-\chi((u+1)z)\right)\left(1+\chi(z^2-4(u+1)z)\right)\left(1+\chi(z^2-4z+4u^2)\right).
\end{split}
\]
%\[
%\begin{split}
%8T_1(u) =&\sum_{z \in \gf_{p^n}\setminus\Omega} (1-\chi(u+1)z)(1+\chi(z^2-4(u+1)z))(1+\chi(z^2-4z+4u^2))\\
%     =&-2+\sum_{z \in \gf_{p^n}}(1-\chi(u+1)z)(1+\chi(z^2-4(u+1)z))(1+\chi(z^2-4z+4u^2))\\
%     =&\sum_{z \in \gf_{p^n}}1-\sum_{z \in \gf_{p^n}}(\chi(u+1)z)+\sum_{z \in \gf_{p^n}}\chi(z^2-4(u+1)z)\\
%     &+\sum_{z \in \gf_{p^n}}\chi(z^2-4z+4u^2)-\sum_{z \in \gf_{p^n}}\chi((u+1)z^3-4(u+1)^2z^2)\\
%     &-\sum_{z \in \gf_{p^n}}\chi((u+1)z^3-4(u+1)z^2+4u^2(u+1)z)\\
%     &+\sum_{z \in \gf_{p^n}}\chi(z^4-4(u+2)z^3+4(u+2)^2z^2-16u^2(u+1)z)\\
%     &-\sum_{z \in \gf_{p^n}}\chi((u+1)z^5-4(u+2)(u+1)z^4+4(u+2)^2(u+1)z^3-16u^2(u+1)^2z^2)-2.
%\end{split}
%\]
By Lemma \ref{charactersumquadratic}, the above expression can be simplified as
\[
\begin{split}
8T_1(u) =& p^n-6-\sum_{z \in \gf_{p^n}}\chi\left((u+1)z^3-4(u+1)z^2+4u^2(u+1)z\right)\\
     &+\sum_{z \in \gf_{p^n}}\chi\left(z^4-4(u+2)z^3+4(u+2)^2z^2-16u^2(u+1)z\right)\\
     &-\sum_{z \in \gf_{p^n}}\chi\left((u+1)z^3-4(u+2)(u+1)z^2+4(u+2)^2(u+1)z-16u^2(u+1)^2\right)\\ 
     =& p^n-6-\Gamma_0(u)-\Gamma_2(u)+\sum_{z \in \gf_{p^n}}\chi\left(z^4-4(u+2)z^3+4(u+2)^2z^2-16u^2(u+1)z\right).    
\end{split}
\]

Next, we will show that
 \[
\begin{split}&\sum_{z \in \gf_{p^n}}\chi\left(z^4-4(u+2)z^3+4(u+2)^2z^2-16u^2(u+1)z\right)\\
=&\sum_{z\in \gf_{p^n}}\chi\left((u+1)^2z^3+(u^2-2u-2)z^2+(1-u^2)z\right)-1\\
=& \Gamma_1(u)-1.\end{split}
\]
Since
\begin{equation*}
\chi\left(z^2-4(u+1)z\right)\chi\left(z^2-4z+4u^2\right)=\chi\left(z^4-4(u+2)z^3+4(u+2)^2z^2-16u^2(u+1)z\right),
\end{equation*} we have
\begin{equation*}
\sum_{z \in \gf_{p^n}}\chi\left(z^4-4(u+2)z^3+4(u+2)^2z^2-16u^2(u+1)z\right) = \sum_{z \in \gf_{p^n}\setminus\Omega}\chi\left(\frac{z^2-4z+4u^2}{z^2-4(u+1)z}\right).
\end{equation*}
Let $t=\frac{z^2-4z+4u^2}{z^2-4(u+1)z}$, thus we have
\begin{equation}\label{t}
(t-1)z^2-[4(u+1)t-4]z-4u^2=0.
\end{equation}
Note that $z=-u$ if and only if $t=1$. When $t\neq1$, \eqref{t} is a quadratic equation in variable $z$, and its discriminant is $16(u+1)^2t^2+16(u^2-2u-2)t+16(1-u^2)$, denoted by $\Delta$. Therefore, we have
\[
\begin{split}
&\sum_{z \in \gf_{p^n}\setminus\Omega}\chi\left(\frac{z^2-4z+4u^2}{z^2-4(u+1)z}\right)\\
&= 1+\sum_{t\neq1}\chi(t)\left(1+\chi(\Delta)\right)\\ 
 &=1+\sum_{t\in \gf_{p^n}}\left(1+\chi\left((u+1)^2t^2+(u^2-2u-2)t+(1-u^2)\right)\right)\chi(t)-2\\ 
   &=\sum_{t\in \gf_{p^n}}\chi(t)+\sum_{t\in \gf_{p^n}}\chi\left((u+1)^2t^3+(u^2-2u-2)t^2+(1-u^2)t\right)-1\\
   &=\sum_{t\in \gf_{p^n}}\chi\left((u+1)^2t^3+(u^2-2u-2)t^2+(1-u^2)t\right)-1\\
    &=\Gamma_1(u)-1.                           
\end{split}
\]This completes the proof.
\hfill$\square$

By Proposition \ref{number}, we have the following corollary.
\begin{corollary}\label{coro1} With the notation of Proposition \ref{number}, let $T(u)=T_1(u)+T_2(u)$ with $u\in\mathcal{U}_1$. Then, we have
	$$8T(u)=2p^n-14+\Gamma_1(u)-\Gamma_2(u)+\Gamma_1(-u)-\Gamma_2(-u),$$
	where $\Gamma_1(u)$ and  $\Gamma_2(u)$ are defined in  \eqref{defofGm}.
\end{corollary}

\begin{proposition}\label{2-uniform}
Let $p^n \equiv 3 \pmod 4 $,  $p^n\geq 7$ and  $f_u(x)$ be the function defined in (\ref{mainfun}). When $u\in\mathcal{U}_{10}\cup \mathcal{U}_{11} $, the differential spectrum of $f_u(x)$ is given by
$$\left[\,\omega_0=(p^n-1)T(u),\,\omega_1=(p^n-1)\left(p^n-2T(u)\right),\,\omega_2=(p^n-1)T(u)\,\right],$$
where $T(u)$  is given in Corollary  \ref{coro1}.
\end{proposition}
{\em Proof:} When $u\in\mathcal{U}_1$, from \eqref{N1} and \eqref{lem3fuNa0}, we have
\begin{equation*}\label{pro6-N1}N_{u,1}(a,b)=\left\{
\begin{array}{ll}
1, &~ \mathrm{if}~b=a^{-1}(1\pm u\chi(a)),\\
0, &~ \mathrm{otherwise},
\end{array} \right.
\end{equation*}
and
\begin{equation*}\label{pro6-N2}
N_{u,2}(a,0)=1.
\end{equation*}
Moreover, by Lemmas \ref{I-IV} and \ref{II-III}, we have 
\begin{equation*}\label{pro6-Nab}
N_{u,2}(a,b)\leq2, ~\forall (a,b)\in\gf_{p^n}^*\times\gf_{p^n}^*,
\end{equation*}
and $N_{u,2}(a,b)=2$ if and only if one of the following conditions holds:

 \noindent(i) $\chi\left(\frac{a(u+1)}{b}\right)=-1$, $\chi\left(a^2b^2-4(u+1)ab\right)=1$ and $\chi\left(a^2b^2-4ab+4u^2\right)=1$;

 \noindent(ii) $\chi\left(\frac{a(u+1)}{b}\right)=1$, $\chi\left(a^2b^2+4(u-1)ab\right)=1$ and $\chi\left(a^2b^2-4ab+4u^2\right)=1$.

When $u\in\mathcal{U}_{10}\cup \mathcal{U}_{11}\subseteq \mathcal{U}_1$, it can be verified that the pairs $(a,b)$ satisfying  $b=a^{-1}(1\pm u)$ don't  meet any of the conditions in Lemmas \ref{I-IV} and \ref{II-III}. Thus, when $u\in\mathcal{U}_{10}\cup \mathcal{U}_{11} $, we have
$N_{u,2}(a,b)=0$ if $b=a^{-1}(1\pm u)$,
and thereby $N_u(a,b)\leq 2$  for any  $(a,b)\in\gf_{p^n}^*\times\gf_{p^n}$. In particular,
$N_u(a,b)=2$ iff $N_{u,2}(a,b)=2$.
The number of pairs $(a,b)$ satisfying the conditions (i) or the conditions  (ii) is equal to $(p^n-1)T(u)$. Thus, in this case the number of pairs $(a,b)\in\mathbb{F}_{p^n}^*\times \mathbb{F}_{p^n}$ such that  $N_u(a,b)=2$ is
 $$(p^n-1)T(u),$$
 which is the value of $\omega_2$ in the differential spectrum. From the identities in \eqref{prop}, the desired result follows.
\hfill$\square$

Next we determine the differential spectrum of $f_u(x)$ for $u\in\mathcal{U}_{12}$.

\begin{proposition}\label{3-uniform} Let $p^n \equiv 3 \pmod 4 $,  $p^n\geq 7$ and  $f_u(x)$ be the function defined in (\ref{mainfun}).  For $u\in\mathcal{U}_{12}$,  the differential spectrum of $f_u(x)$ is given by	
\begin{equation*}
\begin{split}[\,&\omega_0=(p^n-1)\left(T(u)+2\right),\,\omega_1=(p^n-1)\left(p^n-2-2T(u)\right),\\
&\omega_2=(p^n-1)\left(T(u)-2\right),\,\omega_3=2(p^n-1)\,],
\end{split}
\end{equation*}
where $T(u)$  is given in Corollary \ref{coro1}.
\end{proposition}
{\em Proof:}  Following the discussions in the proof of Proposition \ref{2-uniform}, we calculate the differential spectrum in this case. The main difference lies in that when $u\in\mathcal{U}_{12}$, the pairs $(a,b)$ satisfying  $b=a^{-1}(1+u)$ meet the conditions (ii) in the proof of Proposition \ref{2-uniform}, and those satisfying  $b=a^{-1}(1-u)$ fulfill the conditions (i) there. This means that in this case $N_{u,2}(a,b)=2$ iff $b=a^{-1}(1\pm u)$. Thus, when $u\in\mathcal{U}_{12}$, $N_u(a,b)\leq 3$ for any  $(a,b)\in\gf_{p^n}^*\times\gf_{p^n}$. In particular,  $N_u(a,b)=3$ iff $b=a^{-1}(1\pm u)$, and $N_u(a,b)=2$ iff $b\neq a^{-1}(1\pm u)$ and $N_{u,2}(a,b)=2$.  The number of pairs $(a,b)\in \gf_{p^n}^*\times \gf_{p^n}$ such that $N_{u,2}(a,b)=2$ is equal to $(p^n-1)T(u)$, and the number of those satisfying $b=a^{-1}(1\pm u)$ is equal to $2(p^n-1)$. According to the arguments above, we get $\omega_3=2(p^n-1)$ and $\omega_2=(p^n-1)(T(u)-2)$. Then, utilizing the identities in \eqref{prop}, the differential spectrum of $f_u(x)$ in this case is obtained. \hfill$\square$

\begin{remark}
(i) Due to Lemmas \ref{lemforNa0}-\ref{II-IV}, for each given $u\in \gf_{p^n}$, we are able to find the necessary and sufficient conditions on $(a,b)$ such that the derivative equation $\mathbb{D}_af_u(x)=b$ has exactly $i$ solution(s), where $0\leq i\leq \delta(f_u)$. Thus, for each given $u\in \gf_{p^n}$, the problem of determining the differential spectrum of $f_u(x)$ is reduced to that of counting the pairs of $(a,b)\in \gf_{p^n}^*\times \gf_{p^n}$ satisfying the corresponding conditions.

(ii) When $u\in \mathcal{U}_1=\left(\mathcal{U}_{10}\cup\mathcal{U}_{11}\right)\cup \mathcal{U}_{12}$, one has $\delta(f_u(x))\leq 3$ and the relevant conditions are more straightforward to outline. Thus, we can readily
give the differential of $f_u(x)$ in Propositions \ref{2-uniform} and \ref{3-uniform}, which generalize the results in Propositions 6 and 5 of \cite{Xia-Bao2023}, respectively. For $u\in\mathcal{U}_0\setminus \{0,\pm 1\}$, by making use of the results in Proposition \ref{4-uniform-con-1}, the differential spectrum of $f_u(x)$ can also be characterized but with more complicated calculations.
\end{remark}

\section{Conclusion}\label{con-remarks}
In this paper, for  $p^n\equiv~3 \pmod 4$ and $p^n\geq 7$, we conduct a comprehensive investigation on the differential properties of the generalized Ness-Helleseth function  $f_u(x)=ux^{d_{1}}+x^{d_{2}}$, where $d_{1} =  {{p^{n}-1}\over {2}} -1 $ and $d_{2} =p^{n}-2$.  The derivative equation $\mathbb{D}_{a}f_u(x)=b$ of this function was fully characterized according to the coefficient $u$, the input difference $a$ and the output difference $b$ (see Lemmas \ref{lemforNa0}, \ref{I-IV}, \ref{II-III} and \ref{II-IV}). Using the obtained results,  we are able to compute the differential uniformity of  $f_u(x)$ for each $u$. Especially, we determine the differential uniformity of $f_u(x)$ for $u\in \mathcal{U}_0\setminus \{0\}$. This extends the results in \cite{Zeng2007} and \cite{Zha2008}, and also gives the necessary and sufficient conditions for $f_u(x)$ to be APN.

Moreover, we can use the results to calculate the differential spectrum of $f_u(x)$. For $u\in \{0,\pm 1\} $, the differential spectrum of $f_u(x)$ has been explicitly given. For $u\in \mathcal{U}_1$, we express the differential spectrum of $f_u(x)$ in terms of some quadratic character sums of some cubic polynomials. Similarly, for $u\in \mathcal{U}_0\setminus \{0,\pm 1\}$, the differential spectrum of $f_u(x)$ can also be characterized through quadratic character sums, but with more detailed calculations. We leave this problem for the interested readers. The results presented in this paper generalized the results in \cite{Ness-Helleseth-2007} and \cite{Xia-Bao2023} for the ternary case.

\section*{Acknowledgment}
Y. Xia, F. Bao and S. Chen were supported in part by the National Natural
Science Foundation of China under Grants 62171479, 61771021, in part by the Fundamental Research Funds for the Central Universities, South-Central Minzu University under Grant CZT20023, and in part by the Fund for Scientific Research Platforms of South-Central Minzu University under Grant PTZ24004. C. Li and T. Helleseth were supported by the Research Council of Norway under Grant 311646/O7.

\section*{Declarations}

Conflict of interest: The authors declared that they have no conflicts of interest in
connection with the work submitted.


\begin{thebibliography}{99}

\bibitem{Leander2022}C. Beierle and G. Leander, ``New instances of quadratic APN functions,'' \textit{IEEE Trans. Inf. Theory}, vol. 68, no. 1, pp. 670-678, 2022.


\bibitem{Coulter97}R. S. Coulter and R. W. Matthews, ``Planar functions and planes of Lenz-Barlotti class II,'' \textit{Des. Codes Cryptogr.}, vol. 10, no. 2, pp. 167-184, 1997.


\bibitem{Coulter2008} R. S. Coulter and M. Henderson, ``Commutative presemifields and semifields,'' \textit{Adv. Math.}, vol. 217, no. 1, pp. 282-304, 2008.


\bibitem{DO1968}P. Dembowski and T. G. Ostrom, ``Planes of order $n$ with collineation groups of order $n^2$,'' \textit{Math. Zeitschrift}, vol. 103, no. 2, pp. 239-258, 1968.


\bibitem{Dy2006}C. Ding and J. Yuan, ``A family of skew Hadamard difference sets,'' \textit{J. Comb. Theory A}, vol. 113, no. 7, pp. 1526-1535, 2006.

\bibitem{DMMPW2003}H. Dobbertin, D. Mills, E. N. M${\rm\ddot{u}}$ller, A. Pott, and W. Willems, ``APN functions in odd characteristic,'' \textit{Discret. Math.}, vol. 267, no. 1-3, pp. 95-112, 2003.


\bibitem{Ganley1975}M. J. Ganley and E. Spence, ``Relative difference sets and quasiregular collineation groups,'' \textit{J. Comb. Theory A}, vol. 19, no. 2, pp. 134-153, 1975.

\bibitem{Faruk2022}F. G${\rm\ddot{o}}$lo${\rm \breve{g}}$lu, ``Biprojective almost oerfect nonlinear functions,'' \textit{IEEE Trans. Inf. Theory}, vol. 68, no. 7, pp. 4750-4760, 2022.

\bibitem{Faruk2023}F. G${\rm\ddot{o}}$lo${\rm \breve{g}}$lu, ``Classification of $(q, q)$-biprojective APN functions,'' \textit{IEEE Trans. Inf. Theory}, vol. 69, no. 3, pp. 1988-1999, 2023.

\bibitem{HRS1999IT}T. Helleseth, C. Rong, and D. Sandberg, ``New families of almost perfect nonlinear power functions,'' \textit{IEEE Trans. Inf. Theory}, vol. 45. no. 2, pp. 475-485, 1999.

\bibitem{FF} R. Lidl and H. Niederreiter, \emph{Finite Fields} (Encyclopedia of Mathematics and Its Applications), vol. 20. Cambridge, U.K.: Cambridge University Press, 1997.

%\bibitem{Man-Xia2022}Y. Man, Y. Xia, C. Li, and T. Helleseth, ``On the differential properties of the power mapping $x^{ p^m +2 }$," {\it Finite Fields Appl.}, vol. 84, no. 10, pp. 1-22, 2022.

\bibitem{Ness-Helleseth-2007} G. J. Ness and T. Helleseth, ``A new family of ternary almost perfect nonlinear mappings,'' \textit{IEEE Trans. Inf. Theory}, vol. 53, no. 7, pp. 2581-2586, 2007.


\bibitem{Nyberg1991}K. Nyberg, ``Perfect nonlinear S-boxes," in \emph{Advances in Cryptology - EUROCRYPT}, vol. 547. D. W. Davies, Ed. Berlin, Germany: Springer, 1991, pp. 378-386.

\bibitem{Nyberg1994}K. Nyberg, ``Differentially uniform mappings for cryptography," in \emph{Advances in Cryptology - EUROCRYPT}, vol. 765. T. Helleseth, Ed. Berlin, Germany: Springer, 1994, pp. 55-64.


\bibitem{SilveEC} J. H. Silverman, \emph{ The Arithmetic of Elliptic Curves}, 2nd ed. Berlin, Germany: Springer, 2009.

\bibitem{Cyclotomy1967} T. Storer, {\it Cyclotomy and Difference Sets.} Chicago, IL, USA: Markham Publishing Company, 1967.


\bibitem{WLZ2020}Y. Wu, N. Li, and X. Zeng, ``Linear codes from perfect nonlinear functions over finite fields,'' \textit{IEEE Trans. Commun.}, vol. 68, no. 1, pp. 3-11, 2020.

\bibitem{Xia-Bao2023}Y. Xia, F. Bao, S. Chen, C. Li, and T. Helleseth, ``More differential properties of the Ness-Helleseth function,'' \textit{IEEE Trans. Inf. Theory}, vol. 70, no. 8, pp. 6076-6090, 2024.

\bibitem{XTD2022}C. Xiang, C. Tang, and C. Ding, ``Shortened linear codes from APN and PN functions,'' \textit{IEEE Trans. Inf. Theory}, vol. 68, no. 6, pp. 3780-3795, 2022.

%\bibitem{YXLIT}H. Yan, Y. Xia, C. Li, T. Helleseth, M. Xiong, and J. Luo, ``The differential spectrum of the power mapping $x^{p^n-3}$,'' \textit{IEEE Trans. Inf. Theory}, vol. 68, no. 8, pp. 5535 - 5547, 2022.


\bibitem{YCD2006}J. Yuan, C. Carlet, and C. Ding, ``The weight distribution of a class of linear codes from perfect nonlinear functions,'' \textit{IEEE Trans. Inf. Theory}, vol. 52, no. 2, pp. 712-717, 2006.

\bibitem{Zeng2007}X. Zeng, L. Hu, Y. Yang, and W Jiang, ``On the inequivalence of Ness-Helleseth APN functions,'' \textit{IACR Cryptol. ePrint Arch}, https://eprint.iacr.org/2007/379.

\bibitem{Zha2008}Z. Zha, {\it Research on Low Differential Unifomity Functions.} Hunan University, Doctoral Dissertation, 2008.


\bibitem{Zhang2020} X. Zhang, {\it The Differential Spectra and Nonlinearity of Some Power Mappings with Low Uniformity.} South-Central University for Nationalities, Master's theses, DOI:10.27710/d.cnki.gznmc.2020.000413, 2020.



	

\end{thebibliography}
\end{document}